\documentclass[11pt,a4paper]{article}

\usepackage{lineno,hyperref}
\modulolinenumbers[5]
\usepackage{authblk}  
\usepackage[utf8]{inputenc} 
\usepackage[T1]{fontenc} 
\usepackage[ngerman, english]{babel} 
\usepackage{scrlayer-scrpage} 
\usepackage{color} 
\usepackage{graphicx} 
\usepackage{booktabs} 
\usepackage{amssymb,amsmath} 
\usepackage{hyperref}		
\usepackage{graphics} 
\usepackage{caption}  
\usepackage{import}
\usepackage[rgb]{xcolor}
\usepackage{float}
\usepackage{bm}
\usepackage{listings}		
\usepackage{pgf}	 
\usepackage{subcaption} 	
\usepackage{lmodern}
\usepackage{gensymb}
\usepackage{comment}
\usepackage[strings]{underscore}
\usepackage{abstract}
\usepackage{titlesec}

\titleformat{\section}
{\normalfont\bfseries\normalsize}
{\thesection}
{1em}
{}
\titleformat{\subsection}
  {\normalfont\itshape\normalsize}
  {\thesubsection}
  {1em}
  {}

\title{Modelling and simulation of electromechanically coupled dielectric elastomers and myocardial tissue using smoothed finite element methods}

\author[a]{Tan Tran}
\author[b]{Denisa Martonov\'{a}}
\author[a]{Sigrid Leyendecker}

\affil[a]{Friedrich-Alexander-Universität Erlangen-Nürnberg, Institute of Applied Dynamics, Immwahrstraße 1, 91058 Erlangen, Germany}
\affil[b]{Friedrich-Alexander-Universität Erlangen-Nürnberg, Institute of Applied Mechanics, Egerlandstraße 5, 91058 Erlangen, Germany}

\date{}

\begin{document}

\maketitle
\thispagestyle{empty}

\hrule
\begin{abstract}
\noindent Computational modelling offers a cost-effective and time-efficient alternative to experimental studies in biomedical engineering. In cardiac electromechanics, finite element method-based simulations provide valuable insights into diseased tissue behaviour and the development of assistive systems often employing dielectric elastomerxcc actuators (DEAs). However, the use of automatically generated tetrahedral meshes, commonly applied due to geometric complexity of the computational domain, often leads to numerical issues including overly stiff responses and volume locking, particularly in incompressible materials. Smoothed finite element methods (S-FEMs) offer a promising alternative by softening the stiffness matrix through gradient smoothing over defined smoothing domains. This work extends S-FEM formulations to electromechanically coupled problems and compares their performance against standard FEM.~We implement and evaluate four approaches in the Abaqus environment through custom user elements: Standard TET-FEM, face-based S-FEM (FS-FEM), node-based S-FEM (NS-FEM), and the hybrid face/node-based S-FEM (FSNS-FEM). Two benchmark problems for nearly incompressible materials are studied: The electrically induced contraction of an isotropic cube representing a simplified DEA and an anisotropic cube representing a section of the myocardium. Reference solutions are obtained using a higher-order method.~Our results demonstrate that FSNS-FEM provides the best balance between accuracy and computational efficiency, closely matching reference data. NS-FEM produces softer results, which is beneficial for anisotropic materials such as the myocardium but less accurate for isotropic elastomers. FS-FEM and standard FEM consistently exhibit overly stiff behaviour, with pronounced volume locking in the myocardial case. These findings support the potential of S-FEMs, in particular FSNS-FEM, for accurate simulation of coupled electromechanical behaviour in complex biomedical applications.\\
\noindent \textit{keywords:} smoothed finite element method; electromechanical coupling; dielectric elastomer; myocardium; cardiac tissue; volumetric locking; incompressible material; Abaqus\\
\end{abstract}

\hrule


\section{Introduction}
\noindent Cardiovascular diseases (CVDs) represent the leading cause of death worldwide \cite{WHO.2021}. CVDs comprise a broad spectrum of pathological conditions involving the heart and vascular system, with ischaemic heart disease and stroke being the most prevalent manifestations~\cite{Townsend.2022}. At the core of cardiac function lies a coupled electromechanical process: An electrical signal propagates through the cardiac tissue, initiating the contraction of cardiac muscle cells, cardiomyocytes, which in turn drives the mechanical action of blood circulation. This coupling between electrical excitation and mechanical contraction, referred to as excitation-contraction coupling, is fundamental to the normal functioning of myocardial tissue. Accurately capturing this behaviour through computational modelling is essential for advancing our understanding of cardiac mechanics and electrophysiology, and for evaluating the behaviour of biologically inspired materials under coupled field conditions.
\noindent The finite element method (FEM) is widely employed in the simulation of biological tissues because it can accommodate complex geometries, heterogeneous materials, and nonlinear deformation. Particularly in cardiac applications, tetrahedral (TET) meshes are favored due to their adaptability to intricate anatomical shapes, and can be efficiently generated through algorithms such as Delaunay triangulation. However, the use of linear TET elements in nearly incompressible soft tissue often leads to well-known numerical issues such as volumetric locking and overly stiff behaviour~\cite{Liu.2019}. Remedies for these challenges include the use of higher-order elements or significantly refined meshes, both of which increase computational cost.
\noindent An alternative approach is the smoothed finite element method (S-FEM), which combines aspects of mesh-free methods with conventional FEM techniques~\cite{Liu.2007,Liu.2019,Martonova.2021b,Martonova.2023}. S-FEM improves solution accuracy and convergence properties without the need for mesh refinement or higher-order basis functions. In contrast to standard FEM, where strains and stresses are derived element-wise, S-FEM computes smoothed gradients over specifically defined smoothing domains (SDs), which are constructed based on the original mesh topology. Depending on the SD configuration, several S-FEM variants exist. For three-dimensional (3D) problems, commonly used schemes include face-based S-FEM (FS-FEM) introduced by~\cite{NguyenThoi.2009b}, node-based S-FEM (NS-FEM) introduced by~\cite{Liu.2009}, and hybrid schemes such as the face-node selective S-FEM (FSNS-FEM) proposed by~\cite{Jiang.2014}.
\noindent Previous studies have demonstrated the effectiveness of S-FEM for modelling large deformation of soft, anisotropic, and (nearly) incompressible materials, including biological tissues. For instance, FS-FEM has been applied to hyperelastic tissue modelling~\cite{Minh.2014} and mechanical simulations of soft organs~\cite{Mendizabal.2017}. FSNS-FEM and edge-node schemes have also been implemented in models of passive myocardium with complex fibre orientations~\cite{Jiang.2014, Jiang.2015}. Further, different S-FEM techniques have been used in simulation of active cardiac mechanics modelled with a time-dependent, rather then potential-dependent, active stress~\cite{Martonova.2021b, Martonova.2023}.
In summary, all of these applications treat solely the mechanical problem via S-FEM, without accounting for the coupling between electrical and mechanical quantities in the smoothing domains.
Recently, S‑FEM has been successfully extended to electromechanical coupling, but only so far in piezoelectric contexts using cell-based smoothing domains~\cite{Cai.2019, Zheng.2019}.
\noindent 
While several finite element formulations exist for electromechanically coupled dielectric elastomers~\cite{Henann.2013, Wang.2016} and cardiac tissue~\cite{Goktepe.2010, Martonova.2023}, these methods predominantly rely on standard FEM discretisations. To date, no work has demonstrated the application of S-FEMs to such coupled problems involving dielectric elastomers or cardiac tissue and a systematic comparison across different S-FEM variants in this context remains unexplored.
This highlights a research gap and provides the motivation for the current study.
\noindent The goal of this work is to develop and evaluate S-FEM-based formulations for modelling electromechanically coupled problems in soft dielectric and biological materials. Specifically, we develop various electromechnically coupled S-FEMs, including FS-FEM, NS-FEM, and FSNS-FEM using linear TET meshes, and compare their performance against standard TET-FEM. The study focuses on both numerical accuracy and computational efficiency in the context of large-deformation simulations.
\noindent This paper is organised as follows: First, we introduce the continuum mechanics framework, including the strong and weak forms of the governing electromechanical equations. Second, we present the constitutive models for hyperelastic, nearly incompressible dielectric elastomers and incompressible myocardial tissue. Third, we outline the numerical solution strategy, describe the integration of S-FEM into the iterative finite element solver, and summarize implementation details. Finally, two benchmark problems are simulated: an electro-active dielectric elastomer cube representing a simplified DEA and a cubic myocardial segment. Results are compared across the FEM and S-FEM schemes to assess differences in solution quality and computational cost.


\section{Coupled problem formulation}
\noindent  In this section, we state the fundamental equations and continuum mechanical framework to describe the electromechanically coupled problem for the dielectric elastomer and the myocardial tissue. First, we introduce fundamental kinematic quantities to describe the large deformation of soft material. Then, the partial differential equations (PDEs) are stated for the electrical polarisation and mechanical deformation. To later apply the numerical methods, we also introduce the weak forms of the governing PDEs.


\subsection{Kinematics of large deformations}
\label{subsec:fundamentals_kinematics}
\begin{figure}[htp!]
	\centering
	\includegraphics[scale=0.2]{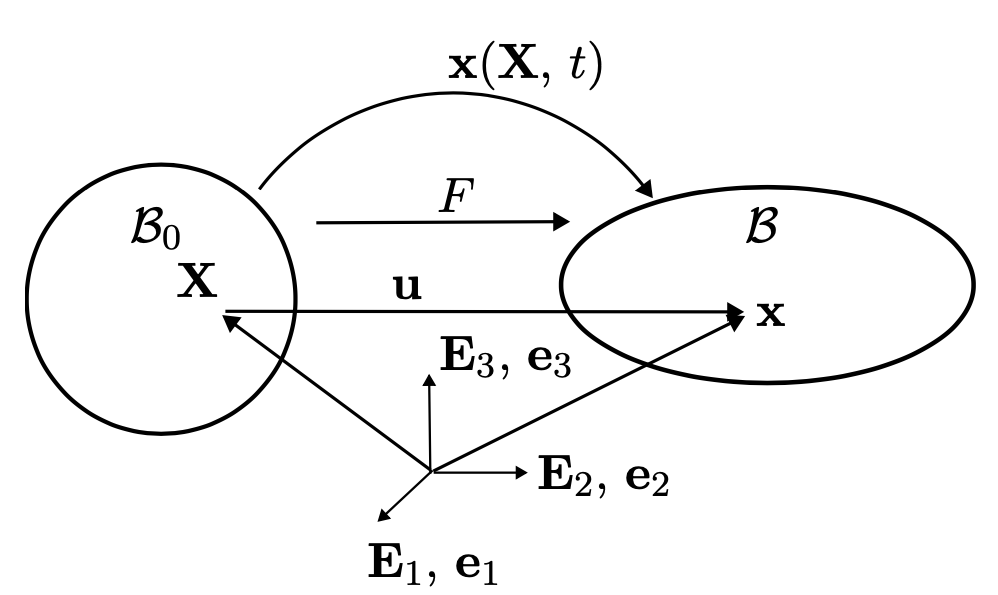}
	\caption{Deformation mapping from the undeformed body $\mathcal{B}_0$ to the deformed body $\mathcal{B}$.}
	\label{fig:fundamentals_mapping}
\end{figure}

\noindent  To describe large deformations, we introduce fundamental quantities based on the finite strain theory which is part of non-linear continuum mechanics. For further information regarding the formulas and theorems, the reader is referred to \cite{Bonet.1997, Wriggers.2008}.\\ 
Let us first consider an undeformed body in the material configuration $\mathcal{B}_0$. Points inside $\mathcal{B}_0$ are described by material coordinates $\mathbf{X}=\mathbf{X}(t_0)$ for a given initial time $t_0$. The deformed body is set in the spatial configuration $\mathcal{B}$ where the position of a point is described by the corresponding spatial coordinates $\mathbf{x}=\mathbf{x}(\mathbf{X},t)$ at a certain time $t \geq t_0$. The spatial coordinates represent unique and continuous differential mappings of the material coordinates from  $\mathcal{B}_0$ to $\mathcal{B}$ with
\begin{equation}
	\mathbf{x}: \mathcal{B}_0 \rightarrow \mathcal{B}, ~ \mathbf{X} \mapsto \mathbf{x}(\mathbf{X},t)
	\label{eq:fundamentals_x} 
\end{equation}
which is known as the Lagrangian description of motion. Both material and spatial coordinates share the same basis. The displacement $\mathbf{u}$ is defined as the difference between the spatial and the material coordinates with
\begin{equation}
	\mathbf{u}(\mathbf{X},t) = \mathbf{x}(\mathbf{X},t) - \mathbf{X}.
	\label{eq:fundamentals_u}
\end{equation}
In non-linear continuum mechanics, the deformation gradient $\mathbf{F}$ is necessary to map tangents between $\mathcal{B}_0$ and $\mathcal{B}$. It is a second order tensor which is computed using the gradient of the deformation mapping with respect to the material coordinates  
\begin{equation}
	\mathbf{F} = \frac{\partial \mathbf{x}}{\partial \mathbf{X}} = \frac{\partial (\mathbf{u+X})}{\partial \mathbf{X}} 
	= \frac{\partial \mathbf{u}}{\partial \mathbf{X}} + \mathbf{I}
	\label{eq:fundamentals_F}
\end{equation}
where $\mathbf{I}$ denotes the second-order identity tensor. The determinant of the deformation gradient is denoted as the Jacobian $J=\text{det}(\mathbf{F})$ and is relevant to describe the volume change during the deformation. The general deformation mapping in non-linear kinematics is illustrated in Figure \ref{fig:fundamentals_mapping}.
There, $\mathbf{E}_i$ denote the material basis vectors and $\mathbf{e}_i$ denote the spatial basis vectors. The following formulas describe the transformation of line, area and volume elements from $\mathcal{B}_0$ to $\mathcal{B}$
\begin{subequations}
\begin{align}
	&d\mathbf{x} = \mathbf{F} d\mathbf{X}, \\
	&\mathbf{n}da = J \mathbf{F}^{-T} \cdot \mathbf{N}dA ~ \text{and} \\
	&dv = J dV \label{eq:fundamentals_vol_transfo}
\end{align}
\label{eq:fundamentals_F_transfo}
\end{subequations}
where $\mathbf{N}$ and $\mathbf{n}$ denote the outward unit normal vectors in $\mathcal{B}_0$ and $\mathcal{B}$ respectively. The existence of $\mathbf{F}^{-T}$ implies that $J$ must be non-zero. Also, the transformation of the volume elements demands $J$ to be non-negative and therefore $J > 0$. To describe the non-linear deformation, we utilize different strain measurements. To account for the geometrical change from $\mathcal{B}_0$ to $\mathcal{B}$, the right Cauchy-Green tensor $\mathbf{C} =  \mathbf{F}^T \mathbf{F}$ is used. Respectively, the left Cauchy-Green tensor $\mathbf{b} = \mathbf{FF}^T$  is used to describe the geometrical change from $\mathcal{B}$ to $\mathcal{B}_0$. In particular, for nearly incompressible materials, it is useful to separate the deformation into a volumetric and isochoric part where the latter does not result in any volume change. The isochoric component $\overline{\mathbf{F}}$ of $\mathbf{F}$ is defined as
\begin{equation}
	\overline{\mathbf{F}} = J^{-\frac{1}{3}} \mathbf{F}.
	\label{eq:fundamentals_F_iso}
\end{equation}
It is evident that $\text{det}(\overline{\mathbf{F}}) = 1$ meaning that there is no volume change. Using the decomposition of $\mathbf{F}$, the isochoric parts of $\mathbf{C}$ and $\mathbf{b}$ are obtained as
\begin{equation}
	\overline{\mathbf{C}} = \overline{\mathbf{F}}^T \overline{\mathbf{F}} = J^{-\frac{2}{3}}  \mathbf{C} ~ \text{and} ~ \overline{\mathbf{b}} = \overline{\mathbf{F}} \overline{\mathbf{F}}^T =  J^{-\frac{2}{3}}  \mathbf{b}.
	\label{eq:fundamentals_strain_iso}
\end{equation}


\subsection{Governing equations of electromechanics}
\label{subsec:fundamentals_strong}
\noindent We now introduce the necessary PDEs to model the electromechanically coupled behaviour. Additionally, essential boundary conditions are defined to complete the strong problem formulation. \\
The mechanical deformation of a body is described by the balance of linear momentum with prescribed Dirichlet and Neumann boundary conditions. Neglecting body forces and inertia forces, the following mechanical boundary value problem (BVP) results
\begin{subequations}
\begin{align}
	&\text{div} (\bm{\sigma}) = \mathbf{0} ~ \text{in} ~ \mathcal{B} \\
	&\mathbf{u} = \overline{\mathbf{u}} ~ \text{on} ~ \mathcal{S}_\mathbf{u} \\
	&\bm{\sigma} \mathbf{n} = \overline{\mathbf{t}} ~ \text{on} ~ \mathcal{S}_\mathbf{t}.
\end{align}
\label{eq:fundamentals_BVP_mech}
\end{subequations}
\noindent In (\ref{eq:fundamentals_BVP_mech}), the Cauchy stress is denoted as $\bm{\sigma}$. Regarding the boundary conditions, the displacement $ \overline{\mathbf{u}} $ and the surface traction force $\overline{\mathbf{t}}$ are prescribed on the complementary boundary domains $\mathcal{S}_{\mathbf{u}}$ and $\mathcal{S}_{\mathbf{t}}$ with $\mathcal{S}_{\mathbf{u}} \cup \mathcal{S}_{\mathbf{t}} = \mathcal{\partial \mathcal{B}}$ and $\mathcal{S}_{\mathbf{u}} \cap \mathcal{S}_{\mathbf{t}} = \emptyset$.\\
Neglecting magnetic effects, the spatial propagation of the electric potential $\varphi$ inside a dielectric elastomer is described by Maxwell's first equation. We adopt the electric BVP for a dielectric elastomer from \cite{Henann.2013} with prescribed boundary conditions as
\begin{subequations}
\begin{align}
	&\text{div} (\mathbf{D}) = 0 ~ \text{in} ~ \mathcal{B}\\
	&\varphi = \overline{\varphi} ~ \text{on} ~ \mathcal{S}_{\varphi}\\
	& -\mathbf{D} \mathbf{n} = \overline{\omega} ~ \text{on} ~ \mathcal{S}_{\omega}.
\end{align}
\label{eq:fundamentals_BVP_el_dielectric}
\end{subequations}
In (\ref{eq:fundamentals_BVP_el_dielectric}), the spatial electric displacement is denoted as $\mathbf{D}$. Regarding the boundary conditions, the electric potential $ \overline{\varphi} $ and the surface charge $\overline{\omega}$ are prescribed on the complementary boundary domains $\mathcal{S}_{\varphi}$ and $\mathcal{S}_{\omega}$ with $\mathcal{S}_{\varphi} \cup \mathcal{S}_{\omega} = \mathcal{\partial \mathcal{B}}$ and $\mathcal{S}_{\varphi} \cap \mathcal{S}_{\omega} = \emptyset$.\\
For the electrical behaviour of the myocardial tissue, we adopt the model stated in \cite{Goktepe.2009}. The spatio-temporal evolution of $\varphi$ inside the myocardium is defined by the following initial BVP
\begin{subequations}
\begin{align}
	&\dot{\varphi} = \text{div} (\mathbf{q}) + I^{\varphi}  ~ \text{in} ~ \mathcal{B} \cup [t_0, \infty) \\
	& \varphi_0 = \varphi(t_0)  ~ \text{in} ~ \mathcal{B} \cup \{t_0 \} \\
	&\varphi = \overline{\varphi} ~ \text{on} ~ \mathcal{S}_{\varphi}  \cup [t_0, \infty)  \\
	&\mathbf{q} \mathbf{n} = \overline{q} ~ \text{on} ~ \mathcal{S}_{q}  \cup [t_0, \infty).
\end{align}
\label{eq:fundamentals_BVP_el_myo}
\end{subequations}
In (\ref{eq:fundamentals_BVP_el_myo}), the diffusion term $\text{div}(\mathbf{q})$ defines the spatial propagation of $\varphi$ depending on the spatial flux vector $\mathbf{q}$. Additionally, the non-linear source term $I^\varphi$ defines the generation and evolution of $\varphi$ over time. For our work, it is computed based on the Aliev-Panfilov model introduced in \cite{Aliev.1996}. Regarding the boundary conditions, the electric potential $ \overline{\varphi} $ and the surface flux $\overline{\mathbf{q}}$ are prescribed on the complementary boundary domains $\mathcal{S}_{\varphi}$ and $\mathcal{S}_{\mathbf{q}}$ with $\mathcal{S}_{\varphi} \cup \mathcal{S}_{\mathbf{q}} = \mathcal{\partial \mathcal{B}}$ and $\mathcal{S}_{\varphi} \cap \mathcal{S}_{\mathbf{q}} = \emptyset$. \\
 In order to later apply the discretisation and numerical solution scheme, we rewrite the PDEs from above into a weak formulation.  The Galerkin method is utilized to derive the weak forms of the PDEs from (\ref{eq:fundamentals_BVP_mech})-(\ref{eq:fundamentals_BVP_el_myo}). Therefore, we introduce the following test functions $\mathbf{v}_1$ and $v_2$ and their respective sets $\mathcal{V}_1$ and $\mathcal{V}_2$ as follows
\begin{subequations}
\begin{align}
	&\mathcal{V}_1 = \{ \mathbf{v}_1 \in \mathcal{H}^1: \mathbf{v}_1 = \mathbf{0} ~ \text{on} ~
	\mathcal{S}_\mathbf{u} \} ~\text{and} \\ 
	&\mathcal{V}_2 = \{ v_2 \in \mathcal{H}^1: v_2 = 0 ~ \text{on} 
	~ \mathcal{S}_{\varphi} \}
	\label{eq:test_func_spaces}
\end{align}
\end{subequations}
where $\mathcal{H}^1$ is the  Sobolev-space. The test functions have the property that they vanish on the Dirichlet boundary domain. The spatial weak form for the mechanical problem in (\ref{eq:fundamentals_BVP_mech}) is given by
\begin{equation}
	\mathcal{G}_\mathbf{u}(\mathbf{u}, \mathbf{v}_1)=-\int_{\mathcal{B}} \frac{\partial \mathbf{v}_1}{\partial \mathbf{x}}:\bm{\sigma} dv
	+ \int_{\mathcal{S}_\mathbf{t}} \mathbf{v}_1 \mathbf{t} da = 0, ~ \forall \mathbf{v}_1 \in \mathcal{V}_1.
	\label{eq:fundamentals_weak_mech}
\end{equation}
The spatial weak form for the electrical dielectric elastomer problem from (\ref{eq:fundamentals_BVP_el_dielectric}) is given by
\begin{equation}
	\mathcal{G}_{\varphi}^{\text{die}}(\varphi, v_2)= \int_{\mathcal{B}} \frac{\partial v_2}{\partial \mathbf{x}} \mathbf{D}  dv 
	- \int_{\mathcal{S}_q} v_2 \overline{\omega}  da = 0, ~ \forall v_2\in \mathcal{V}_2.
	\label{eq:fundamentals_weak_dielectric}
\end{equation}
Both formulations correspond to the weak forms stated in \cite{Henann.2013}. The spatial weak form for the electrical myocardial tissue problem from (\ref{eq:fundamentals_BVP_el_myo}) is derived in \cite{Goktepe.2010} to
\begin{equation}
	\mathcal{G}_{\varphi}^{\text{myo}}(\varphi, v_2)= \int_{\mathcal{B}} [v_2 \dot{\varphi}  +  \frac{\partial v_2}{\partial \mathbf{x}} \mathbf{q}- V_2 I^{\varphi}] dv 
	- \int_{\mathcal{S}_q} v_2 \overline{q}da = 0, ~ \forall v_2\in \mathcal{V}_2.
	\label{eq:fundamentals_weak_myo}
\end{equation}
It can be shown that solutions of the strong formulations in (\ref{eq:fundamentals_BVP_mech})-(\ref{eq:fundamentals_BVP_el_myo}) also satisfy the corresponding weak formulation in (\ref{eq:fundamentals_weak_mech})-(\ref{eq:fundamentals_weak_myo}). The use of the weak formulations poses less continuity requirements on the solution making it suitable for the use in FEM.


\section{Constitutive equations}
\noindent In this section, we summarize the constitutive equations for the dielectric elastomer and the myocardial tissue. These equations relate the solution variables  $\mathbf{u}, \varphi$ to the stress and electrical quantities from (\ref{eq:fundamentals_BVP_mech}) - (\ref{eq:fundamentals_BVP_el_myo}) and therefore characterise the coupling behaviour. 


\subsection{Constitutive equations for the dielectric elastomer}
\label{subsec:material_dielectric}
\noindent Based on Gauss's law, $\mathbf{D}$ from (\ref{eq:fundamentals_BVP_el_dielectric}) is computed as 
\begin{equation}
	\mathbf{D} = \epsilon \mathbf{E}.
	\label{eq:material_D}
\end{equation}
There,  $\mathbf{E}=-\frac{\partial \varphi}{\partial \mathbf{x}}$ denotes the spatial electric field and $\epsilon$ the material-dependent dielectric permittivity. The corresponding material electric field is computed by the pull-back operation $\mathbf{E}_0=\mathbf{F}^{T} \mathbf{E}$ with  $\mathbf{E}_0=-\frac{\partial \varphi}{\partial \mathbf{X}}$. Hence, the polarisation of the dielectric elastomer is also governed by the mechanical deformation.
We further adopt the assumption from \cite{Henann.2013} and neglect the effects of free space surrounding the dielectric for simplistic reasons. Hence, the permittivity of free space $\epsilon_0$ is set to zero. Regarding the mechanics, $\bm{\sigma}$ is decomposed into an active part $\bm{\sigma}^\text{act}$ and passive  part $\bm{\sigma}^\text{pas}$. The electrical polarisation generates $\bm{\sigma}^\text{act}$ based on \cite{Zhao.2007}, while $\bm{\sigma}^\text{pas}$ is purely mechanical. Additionally, we separate $\bm{\sigma}^\text{pas}$ into a volumetric part $\bm{\sigma}^\text{vol}$ and isochoric part $\bm{\sigma}^\text{iso}$ with $\bm{\sigma}^\text{pas}$ =  $\bm{\sigma}^\text{vol}$ +  $\bm{\sigma}^\text{iso}$. For the dielectric elastomer, we define $\bm{\sigma} $ as
\begin{equation}
	\bm{\sigma} = \underbrace{\epsilon [\mathbf{E} \otimes \mathbf{E} - \frac{1}{2} (\mathbf{E} \mathbf{E}) \mathbf{I}]}_{\bm{\sigma}^\text{act}} 
	+\underbrace{\kappa (J-1) \mathbf{I} }_{\bm{\sigma}^\text{vol}}
	+  \underbrace{\frac{\mu}{J} [\overline{\mathbf{b}}-\frac{1}{3} \text{tr}(\overline{\mathbf{b}})\mathbf{I}]}_{\bm{\sigma}^\text{iso}}
	\label{eq:material_stress_dielectric}
\end{equation}
The definition for $\bm{\sigma}^\text{act}$ is adopted from \cite{Zhao.2007} and the definition of $\bm{\sigma}^\text{pas}$ is adopted from \cite{Bonet.1997}. The parameters $\mu$ and $\lambda$ denote the first and second Lam\'{e}'s parameter and are material-dependent. The passive isotropic behaviour is described by the Neo-Hookean model for the nearly incompressible case.


\subsection{Constitutive equations for the myocardium}
\label{subsec:material_myo}
\begin{figure}[htpb!]
	\centering
	\includegraphics[scale=0.2]{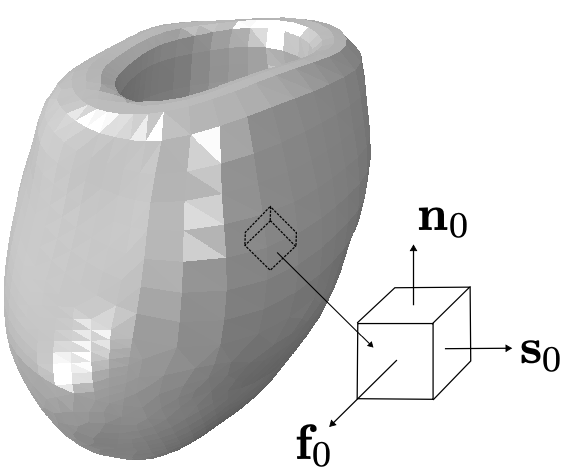}
	\caption{Anisotropic model of the myocardium. The orientation of a single myocardial cell (symbolised by a cube) is defined by an orthonormal $\mathbf{f}_0-\mathbf{s}_0-$$\mathbf{n}_0$-basis. There, $\mathbf{f}_0$ denotes the fibre direction, $\mathbf{s}_0$ the sheet direction and $\mathbf{n}_0$ the normal direction in material configuration.}
	\label{fig:material_left_ventricle}
\end{figure}

\noindent As mentioned in Section \ref{subsec:fundamentals_strong}, the term $I^\varphi$ from (\ref{eq:fundamentals_BVP_el_myo}) is computed based on the the Aliev-Panfilov model from \cite{Aliev.1996}. Unlike in \cite{Goktepe.2010}, the term $I^\varphi$ is not influenced by the mechanical deformation in our work. For further information regarding the explicit computation of $I^\varphi$, the reader is referred to \cite{Aliev.1996, Goktepe.2009}. The flux vector $\mathbf{q}$ is defined in \cite{Goktepe.2009} as
\begin{equation}
	\mathbf{q} = d \frac{\partial \varphi}{\partial \mathbf{x}} = - d  \mathbf{E}
\label{eq:material_q}
\end{equation}
where $d$ is the scalar conductivity. For this work, we assume isotropy for the spatial propagation of $\varphi$. Hence, a scalar representation of the conductivity is sufficient. This assumption is made to reduce the complexity of the numerical implementation. Similar to the dielectric elastomer, we decompose $\bm{\sigma}$ inside the myocardial tissue into $\bm{\sigma}^\text{act}$ and $\bm{\sigma}^\text{pas}$ with an additional subdivision in $\bm{\sigma}^\text{vol}$  and $\bm{\sigma}^\text{iso}$ with $\bm{\sigma}^\text{pas} = \bm{\sigma}^\text{vol}+\bm{\sigma}^\text{iso}$
\begin{equation}
	\bm{\sigma} = \underbrace{ \frac{1}{J} T(\varphi) \mathbf{f} \otimes \mathbf{f}}_{\bm{\sigma}^\text{act}} 
	+  \underbrace{ \kappa(J-1) \mathbf{I}}_{\bm{\sigma}^\text{vol}} 
	+ \underbrace{\overline{\bm{\sigma}} : \mathbb{P}}_{\bm{\sigma}^\text{iso}}
	\label{eq:material_stress_myo}
\end{equation}
with the active fibre tension $T(\varphi)$, the spatial fibre direction $\mathbf{f}=\mathbf{Ff}_0$ and the modified Cauchy stress $\overline{\bm{\sigma}}$. The material directions $\mathbf{f}_0$ and $\mathbf{s}_0$ define the fibre orientation of the myocardial muscle cell as illustrated in Figure \ref{fig:material_left_ventricle}.
To compute $\bm{\sigma}^\text{act}$, $T(\varphi)$ is evaluated based on the given ordinary differential equation in \cite{Nash.2004}
\begin{equation}
	\dot{T} = a(\varphi) [k_T (\varphi-\varphi_r) - T]
	\label{eq:material_fibre_tension}
\end{equation}
with the control parameter $k_T$, the resting potential $\varphi_r$ and the switch function $a(\varphi)$. The parameter $k_T$ controls the amplitude of $\bm{\sigma}^\text{act}$. The switch function is a smoothed form of the Heaviside step function. It is explicitly defined by
\begin{equation}
	a(\varphi) = a_0 + (a_\infty - a_0) \text{exp}[-\text{exp}(-\xi(\varphi-\overline{\varphi}))]
	\label{eq:material_switch_func}
\end{equation}
where $\overline{\varphi}$ is the phase shift and $a_0$ and $a_\infty$ are limiting parameters. In contrast to \cite{Goktepe.2010}, the following inequality $a_0 > a_\infty$ must hold. This relation is proposed in \cite{Eriksson.2013} in order to achieve a temporal delay between the peak of $\bm{\sigma}^\text{act}$ and the amplitude of $\varphi$. The function approaches $a_\infty$ for $\varphi < \overline{\varphi}$ and $a_0$ for $\varphi > \overline{\varphi}$. The parameter $\xi$ controls the transition rate. A backward Euler scheme is used to compute $T(\varphi)$ iteratively.\\
In (\ref{eq:material_stress_myo}), $\bm{\sigma}^\text{vol}$ is defined by the material-dependent bulk modulus $\kappa$ and $\bm{\sigma}^\text{iso}$ is computed based on the Holzapfel-Ogden model. This model was introduced in \cite{Holzapfel.2000, Holzapfel.2009} and expanded in \cite{Goktepe.2011} to account for nearly incompressible material. To formulate $\overline{\bm{\sigma}}$, we first state the Holzapfel-Ogden strain energy function $\Psi$ given in \cite{Goktepe.2011}
\begin{subequations}
\begin{align}
	&\Psi =  \Psi^{\text{iso}} +\Psi^f +\Psi^s + \Psi^{fs} ~ \text{with} \nonumber \\
	&\Psi^{\text{iso}} = \frac{a}{2b} \text{exp}[b(\overline{I}_{\text{iso}} - 3)],\\
	&\Psi^{\text{f}} = \frac{a_{f}}{2b_{f}} [\text{exp}[b_{f} (\overline{I}_{{f}} - 1)^2]-1], \\
	&\Psi^{\text{s}} = \frac{a_{s}}{2b_{s}} [\text{exp}[b_{s} (\overline{I}_{{s}} - 1)^2]-1] ~ \text{and} \\
	&\Psi^{fs} = \frac{a_{fs}}{2b_{fs}} [\text{exp}[b_{fs} (\overline{I}_{fs})^2]-1]. 
\end{align} 
\label{eq:material_HO_energy}
\end{subequations}

The energy is formulated in terms of the isochoric invariants $\overline{I}_{\text{iso}}$,  $\overline{I}_{f}$,  $\overline{I}_{s}$ and  $\overline{I}_{fs}$ and independent material parameters  $a$, $b$, $a_f$, $b_f$, $a_s$, $b_s$, $a_{fs}$ and $b_{fs}$. The invariants are defined by
\begin{subequations}
    \begin{align}
	&\overline{I}_{\text{iso}} = \text{tr}(\overline{\mathbf{b}}), ~ \overline{I}_{f} =  \text{tr}(\overline{\mathbf{f}} \otimes \overline{\mathbf{f}}), \\ 
	&\overline{I}_{s} =\text{tr}(\overline{\mathbf{s}} \otimes \overline{\mathbf{s}}) ~ \text{and} ~
	~ \overline{I}_{fs} = \text{tr}[(\overline{\mathbf{f}} \otimes \overline{\mathbf{s}})^\text{sym}]
\end{align}
\label{eq:material_HO_invariants}
\end{subequations}
with the spatial isochoric directions $\overline{\mathbf{f}} = \overline{\mathbf{F}} \mathbf{f}_0$ and $\overline{\mathbf{s}} = \overline{\mathbf{F}} \mathbf{s}_0$ and the isochoric left Cauchy-Green tensor $\overline{\mathbf{b}}$ from (\ref{eq:fundamentals_strain_iso}). According to \cite{Goktepe.2011}, the explicit expression of $ \overline{\bm{\sigma}}$ is
\begin{align}
	 \overline{\bm{\sigma}} &= \frac{1}{J} [2 \frac{\partial\Psi^{\text{iso}}}{\partial \overline{I}_{\text{iso}}} \overline{\mathbf{b}} + 2 \frac{\partial\Psi^{\text{f}}}{\partial \overline{I}_f} \overline{\mathbf{f}} \otimes \overline{\mathbf{f}} 
	  + 2 \frac{\partial\Psi^{\text{s}}}{\partial \overline{I}_s} \overline{\mathbf{s}} \otimes \overline{\mathbf{s}} \nonumber \\
	  &+  \frac{\partial\Psi^{\text{fs}}}{\partial \overline{I}_{fs}} (\overline{\mathbf{f}} \otimes \overline{\mathbf{f}} + \overline{\mathbf{s}} \otimes \overline{\mathbf{s}})].
	  \label{eq:material_HO_modified_stress}
\end{align}
$\bm{\sigma}^\text{iso}$ is then computed by a double contraction between $ \overline{\bm{\sigma}}$ and the isochoric fourth-order projection tensor $ \mathbb{P} =  \mathbb{I}^\text{sym} - \frac{1}{3} \mathbf{I} \otimes \mathbf{I}$ in which $\mathbb{I}^\text{sym}$ is the symmetric fourth-order identity tensor defined by $\mathbb{I}^\text{sym}_{ijkl} = \frac{1}{2} (\delta_{ik} \delta_{jl} + \delta_{il} \delta_{jk})$. 


\section{Numerical solution procedure using S-FEM}
\noindent In this section, we summarize the numerical approximations of the governing equations and solution variables. First, some fundamentals of the FEM theory are introduced which serve as basics for S-FEM.  Then, the approximation scheme using S-FEM is presented. To implement the numerical solution scheme, the discretisation of the spatial weak formulations is described. We mainly adopt the descriptions from \cite{Wriggers.2008} regarding the non-linear FEM and the descriptions from \cite{Liu.2010} regarding the S-FEM formulation.


\subsection{Finite element method}
\label{subsec:numerics_fem}
\noindent In standard FEM, the original geometry $\mathcal{B}$ is subdivided into so-called finite elements $\mathcal{B}^e$. The elements are connected to each other through nodes and form an approximation geometry $\mathcal{B}^h$, such that $\mathcal{B} \approx \mathcal{B}^h = \bigcup_{e=1}^{N_{el}} \mathcal{B}^e$. There, $N_{el}$ is the total number of $\mathcal{B}^e$. The element-wise solution variables $\mathbf{u}^e$, $\varphi^e$ are approximated on each $\mathcal{B}^e$ as 
\begin{align}
	&\mathbf{u}^e = \sum_{a=1}^{n_n}  \mathbf{u}^a N^a ~ \text{and} \\
	&\varphi^e = \sum_{a=1}^{n_n}  \varphi^a N^a 
	\label{eq:numerics_element_solution}
\end{align}
where $n_n$ is the number of element nodes and $N^a$ is the nodal shape function. The index $a$ denotes the local node number. Each quantity is evaluated at $a$ and weighted by the corresponding $N^a$. Following the Galerkin method, the element-wise test functions $\mathbf{v}_1^e$, $v_2^e$ can also be approximated by the same shape functions as 
\begin{align}
&\mathbf{v}_1^e = \sum_{a=1}^{n_n}  \mathbf{v}_1^a N^a  ~ \text{and} \\
&v_2 ^e= \sum_{a=1}^{n_n}  v_2^a N^a.
\end{align}
For the discretisation of the weak forms from Section \ref{subsec:fundamentals_strong}, the test function gradients also need to be computed element-wise. Only the shape functions are affected by the gradient operator. Therefore, the test function gradients result in 
\begin{align}
&\frac{\partial \mathbf{v}_1^e}{\partial \mathbf{x}} = \sum_{a=1}^{n_n}  \mathbf{v}_1^a \otimes \frac{\partial N^a}{\partial \mathbf{x}}  ~ \text{and} \\
&\frac{\partial v_2^e}{\partial \mathbf{x}} = \sum_{a=1}^{n_n}  v_2^a  \frac{\partial N^a}{\partial \mathbf{x}}.
\end{align}
Additionally, we can approximate the element-wise deformation gradient $\mathbf{F}^e$ and electric field $\mathbf{E}^e$ as 
\begin{align}
	&\mathbf{F}^e = \sum_{a=1}^{n_n} \mathbf{u}^a \otimes \frac{\partial N^a}{\partial \mathbf{X}} + \mathbf{I} ~ \text{and} \\
   &\mathbf{E}^e= -\sum_{a=1}^{n_n} \varphi^a (\mathbf{F}^e)^{-T} \frac{\partial N^a}{\partial \mathbf{X}}.
	\label{eq:numerics_element_grad}
\end{align}


\subsection{Smoothed finite element method}
\label{subsec:numerics_sfem}
\begin{figure}[h!]
	\centering
	\includegraphics[scale=0.2]{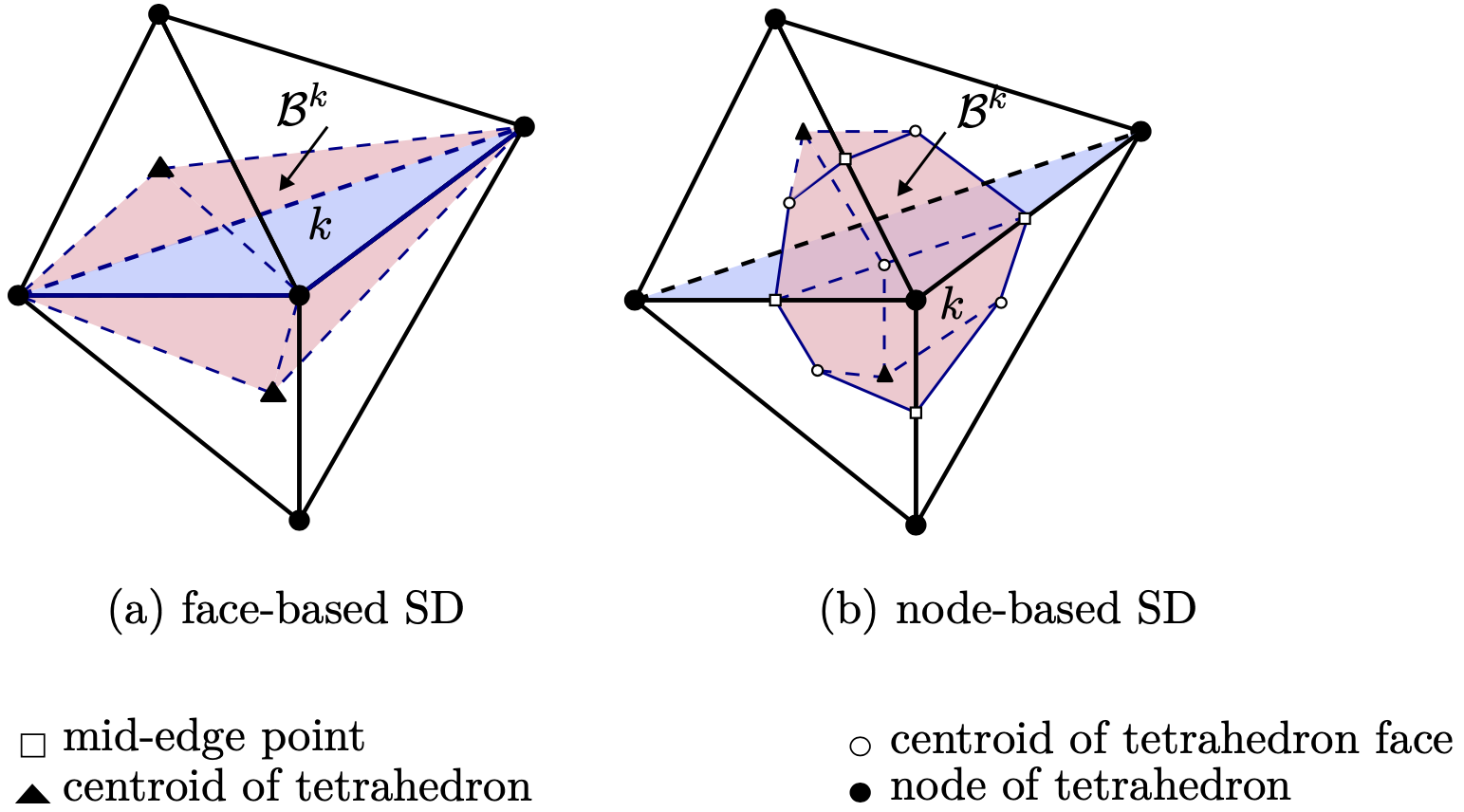}
	\caption{SDs on an initial 3D TET mesh where $\mathcal{B}^k$ for FS-FEM is constructed around a connecting face $k$ in (a) and for NS-FEM, $\mathcal{B}^k$ is constructed around a connecting node $k$ in (b) (adopted from \cite{Martonova.2023}).}
	\label{fig:numerics_SD}
\end{figure}

\noindent The basic concept behind S-FEM is to apply a smoothing operation on the gradients over defined SDs, according to \cite{Liu.2007}. The spatial body $\mathcal{B}$ is subdivided into non-overlapping cells such that $\mathcal{B} \approx \bigcup_{k=1}^{N_s} \mathcal{B}^k$.
There, $\mathcal{B}^k$ is a SD and $N_s$ is the total number of SDs. The resulting SD mesh is over-layered on an initial FEM mesh. This work focuses on two main smoothing methods in 3D, FS-FEM presented in \cite{NguyenThoi.2009b} and NS-FEM presented in \cite{NguyenThoi.2010}. In FS-FEM, the smoothing domain is created around the connecting faces of neighbouring elements. First, the connecting face between two elements is identified. Then, the nodes of each face are connected with the respective element centroids to form the face-based SDs. For elements on the boundary, the nodes on the boundary face are connected to the centroid of the boundary element. The total number of SDs in FS-FEM is equal to the number of faces in the FEM mesh. 
In NS-FEM, each SD is created around the connecting node of adjacent elements. For each node, the connecting elements are identified. Then, the mid-points of the edges connected to the node, the connected element centroids and the centroids of the connected faces together with the node itself form the node-based SD. The volume of the node-based SD depends on the number of adjacent elements. The total number of node-based SDs is equal to the number of nodes in the FEM mesh. In this work, the SDs are constructed based on four node tetrahedral elements. In Figure \ref{fig:numerics_SD}, the resulting SDs are depicted for FS-FEM and NS-FEM.
In \cite{Liu.2007}, the following smoothing function $\Phi(\mathbf{X})$ is introduced
\begin{equation}
	\Phi(\mathbf{X}) = \begin{cases} \frac{1}{V^k} ~ \text{if} ~ \mathbf{X} \in \mathcal{B}^k \\
		0 ~ \text{else} \end{cases}
	\label{eq:numerics_sfem_smoothing}
\end{equation}
where $V^k$ is the volume of the smoothing domain with $V^k = \int_{\mathcal{B}^k} dV$. Additionally, $\Phi(\mathbf{X})$ also satisfies the positivity and unity conditions 
\begin{equation}
	\Phi(\mathbf{X}) \geq 0 ~ \text{and} ~ \int_{\mathcal{B}^k} \Phi(\mathbf{X}) dV = 1.
\end{equation}
The original idea behind S-FEM is to apply $\Phi(\mathbf{X})$ to the element-wise strains connected to the respective SD leading to a smoothed deformation gradient. Using (\ref{eq:numerics_sfem_smoothing}), the smoothed deformation gradient $\mathbf{F}^k$ can be derived from
\begin{equation}
	\mathbf{F}^k = \int_{\mathcal{B}^k} \frac{\partial \mathbf{u}}{\partial \mathbf{X}} \Phi(\mathbf{X}) dV + \mathbf{I} = \frac{1}{V^k}  \int_{\mathcal{B}^k} \frac{\partial \mathbf{u}}{\partial \mathbf{X}} dV + \mathbf{I}.
	\label{eq:numerics_sfem_F0}
\end{equation}
The SDs around the faces or nodes consist of contributions from the adjacent elements. Therefore, (\ref{eq:numerics_sfem_F0}) can be rewritten into
\begin{equation}
	\mathbf{F}^k =  \frac{1}{4V^k} \sum_{e=1}^{n_k} \frac{\partial \mathbf{u}}{\partial \mathbf{X}} V^e + \mathbf{I}
	\label{eq:numerics_sfem_F1}
\end{equation}
where $n_k$ is the number of adjacent tetrahedrons $\mathcal{B}^e$ around node $k$ or face $k$ and $V^e$ is the volume of each $\mathcal{B}^e$. With further transformations and based on \cite{Martonova.2023}, (\ref{eq:numerics_sfem_F1}) results in 
\begin{equation}
		\mathbf{F}^k = \frac{1}{4V^k} \sum_{e=1}^{n_k} \underbrace{{\sum_{a=1}^{n_n} (\mathbf{u}^a \otimes \frac{\partial N^a}{\partial \mathbf{X}}} + \mathbf{I})}_{\mathbf{F}^e} V^e
		\label{eq:numerics_sfem_F}
\end{equation}
where $n_n$=4 because of the use of tetrahedral elements. Due to the electromechanical coupling, the gradient smoothing operation also affects the electric field $\mathbf{E}$ which results in a smoothed spatial electric field $\mathbf{E}^k$. Adopting the same steps used for the derivation of  (\ref{eq:numerics_sfem_F}), the explicit approximation of $\mathbf{E}^k$ results in 
\begin{equation}
	  \mathbf{E}^k=- \frac{1}{4V^k} \sum_{e=1}^{n_k} \underbrace{\sum_{a=1}^{n_n} \varphi^a (\mathbf{F}^e)^{-T} \frac{\partial N^a}{\partial \mathbf{X}}}_{-\mathbf{E}^e} V^e .
	\label{eq:smoothed_electric_field}
\end{equation}
Additionally to the gradient smoothing, the solution variables are also affected by the smoothing operation. The smoothed solutions  $\mathbf{u}^k$, $\varphi^k$ correspond to the solution on the SD. Applying $\Phi(\mathbf{X})$ on  $\mathbf{u}$ and $\varphi$ results in 
\begin{align}
	&\mathbf{u}^k  = \frac{1}{4V^k} \sum_{e=1}^{n_k} \underbrace{\sum_{a=1}^{4} \mathbf{u}^a N^a}_{\mathbf{u}^e} V^e \label{eq:smoothed_u} ~ \text{and} \\
	&  \varphi^k=  \frac{1}{4V^k} \sum_{e=1}^{n_k} \underbrace{\sum_{a=1}^{4} \varphi^a N^a}_{\varphi^e}  V^e  \label{eq:smoothed_phi}. 
\end{align}
To complete the approximations, we also state the smoothed test functions $\mathbf{v}_1^k$, $v_2^k$ as
\begin{align}
&\mathbf{v}_1^k = \frac{1}{4V^k} \sum_{e=1}^{n_k} \underbrace{ \sum_{a=1}^{n_n}  \mathbf{v}_1^a N^a}_{\mathbf{v}_1^e} V^e ~ \text{and} \\
&v_2^k = \frac{1}{4V^k} \sum_{e=1}^{n_k} \underbrace{ \sum_{a=1}^{n_n}  v_2^a N^a}_{v_2^e} V^e.
\end{align}
The smoothed test function gradients then result in 
\begin{align}
&\frac{\partial \mathbf{v}_1^k}{\partial \mathbf{x}} =  \frac{1}{4V^k} \sum_{e=1}^{n_k} \underbrace{\sum_{a=1}^{n_n}  \mathbf{v}_1^a \otimes \frac{\partial N^a}{\partial \mathbf{x}}}_{\partial \mathbf{v}_1^e / \partial \mathbf{x}} V^e
~ \text{and} \label{eq:numerics_dv1_smoothed}\\ 
&\frac{\partial v_2^k}{\partial \mathbf{x}} = \frac{1}{4V^k} \sum_{e=1}^{n_k} \underbrace{\sum_{a=1}^{n_n}  v_2^a \frac{\partial N^a}{\partial \mathbf{x}}}_{\partial v_2^e / \partial \mathbf{x}} V^e. \label{eq:numerics_dv2_smoothed}
\end{align}
We observe, that the smoothed quantities are just the averaged values of the corresponding element-wise quantities over all elements that are connected to the SD.
\subsection{Spatial discretisation and implementation}
\label{subsec:numerics_dicretisation}
\begin{figure}[h!]
	\centering
	\includegraphics[scale=0.2]{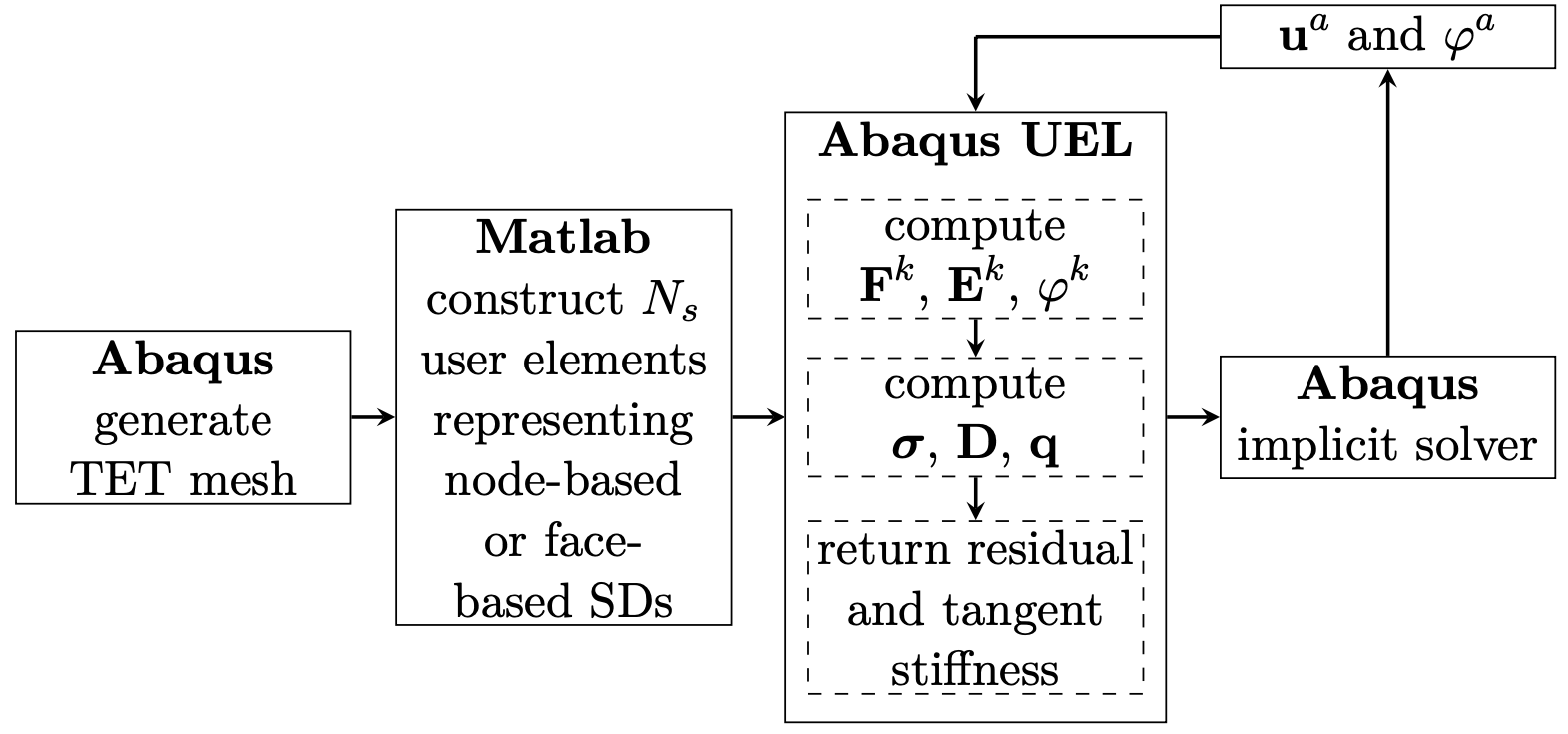}
	\caption{Generalised workflow of the major steps to implement S-FEM into Abaqus for electromechanically coupled problems.}
	\label{fig:numerics_workflow}
\end{figure}

\noindent Inserting the numerical approximations from above into (\ref{eq:fundamentals_weak_mech})-(\ref{eq:fundamentals_weak_myo}), results in the discretised weak forms which are given by local residuals for each element or SD. In S-FEM, the solution-dependent variables are evaluated inside the SD rather than inside the element. 
Besides the different integration domains, the discretisation procedure for S-FEM and standard FEM remains similar. In the following, we only state the local S-FEM residuals. First, the mechanical problem defined by the weak form in (\ref{eq:fundamentals_weak_mech}) is discretised. We adopt the assumptions from \cite{Henann.2013} and neglect the surface traction force for this work. Therefore, the local residual $\mathbf{R}^a_\mathbf{u}$  results in
\begin{equation}
	\mathbf{R}^a_\mathbf{u} = -\int_{\mathcal{B}^k} \nabla N^a \bm{\sigma} dv = \mathbf{0}.
	\label{eq:numerics_Ru}
\end{equation}
Depending on the chosen material model, the smoothed Cauchy stress $\bm{\sigma} = \bm{\sigma}(\mathbf{F}^k, \mathbf{E}^k, \varphi^k)$ is either derived from (\ref{eq:material_stress_dielectric}) or (\ref{eq:material_stress_myo}). For the sake of compactness, we denote the spatial shape function gradient as $\nabla N^a=\frac{\partial N^a}{\partial \mathbf{x}}$ from now on. \\
The local residual for the dielectric elastomer problem $R^a_{\varphi, \text{die}}$  is derived from (\ref{eq:fundamentals_weak_dielectric}). We again adopt the assumption from \cite{Henann.2013} and neglect the surface charge. Therefore, $R^a_{\varphi, \text{die}}$ results in 
\begin{equation}
	R^a_{\varphi, \text{die}}= \int_{\mathcal{B}^k} \nabla N^a \mathbf{D} dv = 0.
	\label{eq:numerics_Rphi_die}
\end{equation}
The smoothed electric displacement $\mathbf{D} = \mathbf{D} (\mathbf{F}^k, \mathbf{E}^k)$ is computed using (\ref{eq:material_D}). For the myocardial tissue, the weak form from (\ref{eq:fundamentals_weak_myo}) is discretised into the local residual $R^a_{\varphi, \text{myo}}$. Based on the FEM discretisation in \cite{Goktepe.2010} and neglecting the surface flux, $R^a_{\varphi, \text{myo}}$ results in 
\begin{equation}
	R^a_{\varphi, \text{myo}} = \int_{\mathcal{B} ^k} [N^a \frac{\varphi^k - \varphi^k_n}{\Delta t} + \nabla N^a \mathbf{q} dv - N^a I^{\varphi}] dv = 0.
	\label{eq:numerics_Rphi_myo}
\end{equation}
In (\ref{eq:numerics_Rphi_myo}), the time derivative of $\varphi$ is approximated by the finite difference $\frac{\varphi^e - \varphi^e_n}{\Delta t}$ where $n$ denotes the previous time and $\Delta t$ the time increment. The smoothed flux $\mathbf{q}=\mathbf{q}(\mathbf{F}^k, \mathbf{E}^k)$ is computed using (\ref{eq:material_q}) and the smoothed source term $I^{\varphi}=I^\varphi(\varphi^k)$ is computed using the Aliev-Panfilov model employed in \cite{Goktepe.2009}. For each element, a local residual $\mathbf{R}^a = (\mathbf{R}^a_\mathbf{u}, R^a_\varphi)^T$ is constructed consisting of the mechanical part $\mathbf{R}^a_\mathbf{u}$ and the electrical part $R^a_\varphi$. Depending on the material, either (\ref{eq:numerics_Rphi_die}) or (\ref{eq:numerics_Rphi_myo}) are used for $R^a_\varphi$. To solve the resulting non-linear system of equations in $\mathbf{R}^a$, the Newton-Raphson method is employed. For the general theory behind this numerical scheme, the reader is referred to \cite{Wriggers.2008}. A more detailed description of the resulting tangent stiffness matrices is presented in Appendix \ref{subsec:appendix_die} and \ref{subsec:appendix_myo} at the end of this paper. \\
In this work, we implemented the numerical framework for standard FEM, FS-FEM, NS-FEM and FSNS-FEM into Abaqus through customized user element subroutines (UELs) . A Matlab script was used to create a user-defined mesh based on an initital TET mesh. The individual user elements of this mesh represent the face-based or node-based SDs. For FSNS-FEM, a node-based SD mesh was overlaid on a face-based SD mesh. The UEL computes and returns the right-hand-side and tangent stiffness matrix of the linearised local residual $\mathbf{R}^a$ for each user element. For further information regarding the creation and structure of an UEL, the reader is referred to the Abaqus documentation \cite{DassaultSystems.2016}. After the UEL computes and returns the local quantities, the solver then assembles them into a global system of equations and computes the node-wise solutions. The workflow is schematically displayed in Figure \ref{fig:numerics_workflow}.
We adopted the UEL code structure from \cite{Wang.2016} and implemented the computation of the smoothed kinematic and constitutive quantities based on \cite{Martonova.2023} for the dielectric elastomer and the myocardial tissue. In Table \ref{tab:numerics_overview_die}, we summarise the numerical methods used for the dielectric elastomer problem. For each numerical method, the table identifies on which element or SD, $\bm{\sigma}^\text{act}$, $\bm{\sigma}^\text{vol}$, $\bm{\sigma}^\text{iso}$ and $\mathbf{D}$ are computed to construct the local residual. The corresponding tangent stiffness matrix is also computed on the respective element or SD.
\begin{table}[h!]
\centering
\begin{tabular}{c  c  c c}
\hline
method/discretisation & TET& face-based SD & node-based SD \\
\hline
FEM & $\bm{\sigma} $, $\mathbf{D}$& ~ & ~ \\
FS-FEM & ~ & $\bm{\sigma} $, $\mathbf{D}$ & ~ \\
NS-FEM & ~ & ~ & $\bm{\sigma}$, $\mathbf{D}$  \\
FSNS-FEM & ~ & $\bm{\sigma}^\text{act}$, $\bm{\sigma}^\text{iso}$, $\mathbf{D}$ & $\bm{\sigma}^\text{vol}$ \\
\end{tabular}
\caption{Numerical methods
used for the dielectric elastomer problem. For each S-FEM, $\bm{\sigma}$ and $\mathbf{D}$ are either being evaluated inside the face-based SDs or node-based SDs or both. For standard FEM, $\bm{\sigma}$ and $\mathbf{D}$ are either being evaluated inside the TET elements.}
\label{tab:numerics_overview_die}
\end{table}
In the same manner, we summarise the numerical methods for the myocardial tissue problem in Table \ref{tab:numerics_overview_myo}. In contrast to the proposed discretisation in (\ref{eq:numerics_Rphi_myo}), $\dot{\varphi}$ and $I^{\varphi}$ are not computed inside the SD for the S-FEMs but on the TET mesh. This work only observes the smoothing effects on the electromechanically coupled variables, namely $\bm{\sigma}(\mathbf{F}^k, \varphi^k)$, $\mathbf{D}(\mathbf{F}^k, \mathbf{E}^k)$ and $\mathbf{q}(\mathbf{F}^k, \mathbf{E}^k)$. For that reason and to ensure comparability between the different methods, the temporal evolution and source are evaluated on the same TET mesh, see Table \ref{tab:numerics_overview_myo}.
\begin{table}[h!]
\centering
\begin{tabular}{c  c  c c}
\hline
method/discretisation & TET& face-based SD & node-based SD \\
\hline
FEM & $\bm{\sigma} $, $\dot{\varphi}$, $\mathbf{q}$, $I^{\varphi}$& ~ & ~ \\
FS-FEM & $\dot{\varphi}$, $I^\varphi$ & $\bm{\sigma}$, $\mathbf{q}$ & ~ \\
NS-FEM & $\dot{\varphi}$, $I^\varphi$  & ~ & $\bm{\sigma}$, $\mathbf{q}$  \\
FSNS-FEM & $\dot{\varphi}$, $I^\varphi$  & $\bm{\sigma}^\text{act}$, $\bm{\sigma}^\text{iso}$, $\mathbf{q}$ & $\bm{\sigma}^\text{vol}$ \\
\end{tabular}
\caption{Numerical methods
used for the myocardial contraction. For each S-FEM, $\bm{\sigma}$ and $\mathbf{q}$ are either being evaluated  inside the face-based SDs or node-based SDs or both. For standard FEM, $\bm{\sigma}$ and $\mathbf{q}$ are either being evaluated inside the TET elements. For each method, $\dot{\varphi}$ and $I^\varphi$ are evaluated inside the TET elements.} 
\label{tab:numerics_overview_myo}
\end{table}

\section{Numerical examples}
\noindent In this section, the implementations of the numerical methods are tested and evaluated with respect to their accuracy and simulation time. A cube model representing a simplified DEA and a section of the myocardium serves as benchmark.


\subsection{Coupled DEA cube}
\label{subsec:examples_die}
\begin{figure}[h!]
	\centering                              
	\includegraphics[scale=0.175]{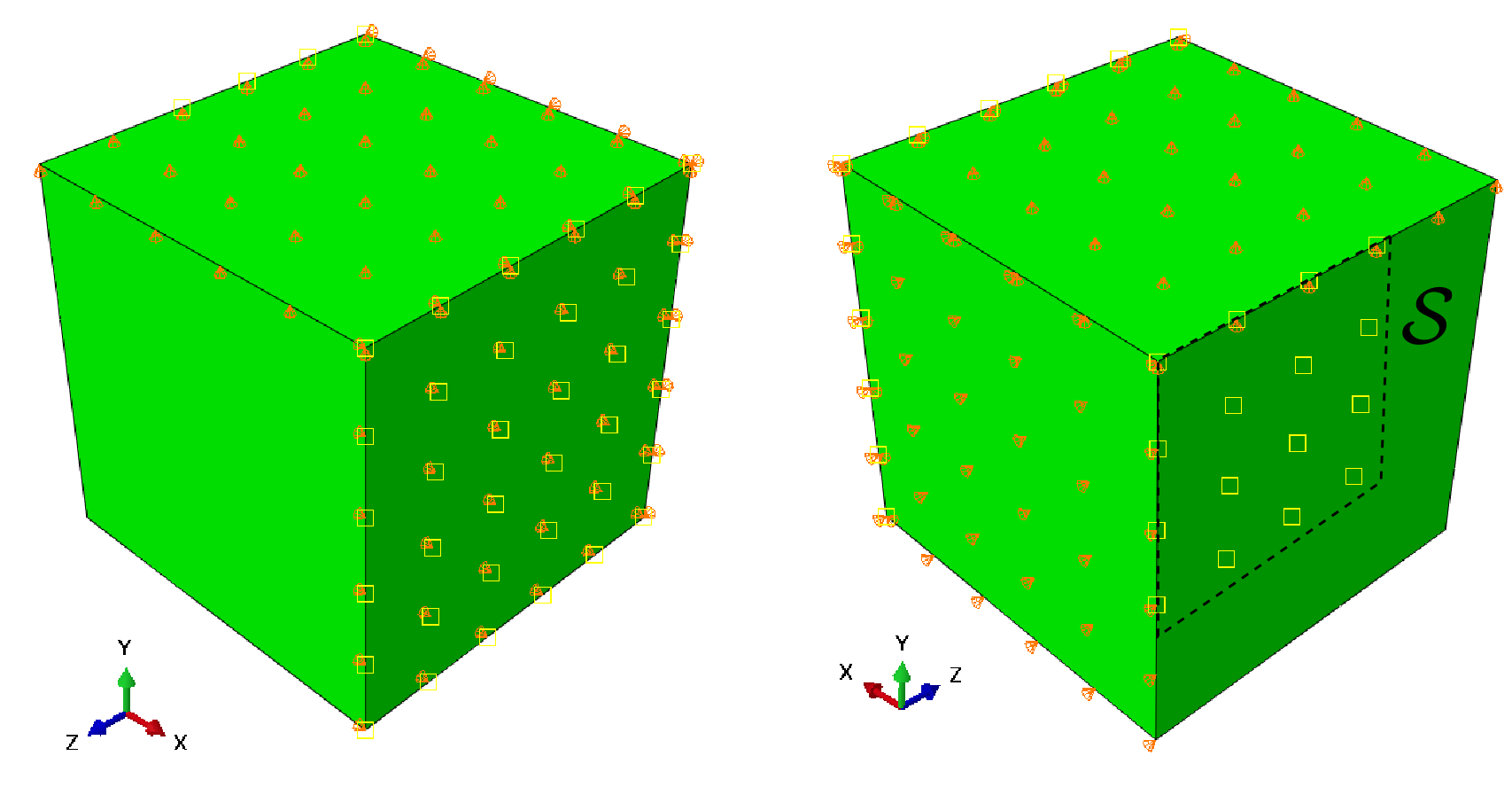}
	\caption{Location of the mechanical (orange cones) and electrical (yellow squares) boundary conditions for the DEA cube simulation. The displacement of the nodes on the excited surface $\mathcal{S}$ are later evaluated.}     
	\label{fig:examples_die_bc}                     
\end{figure}

\begin{figure}[h!]
	\centering                              
	\includegraphics[scale=0.2]{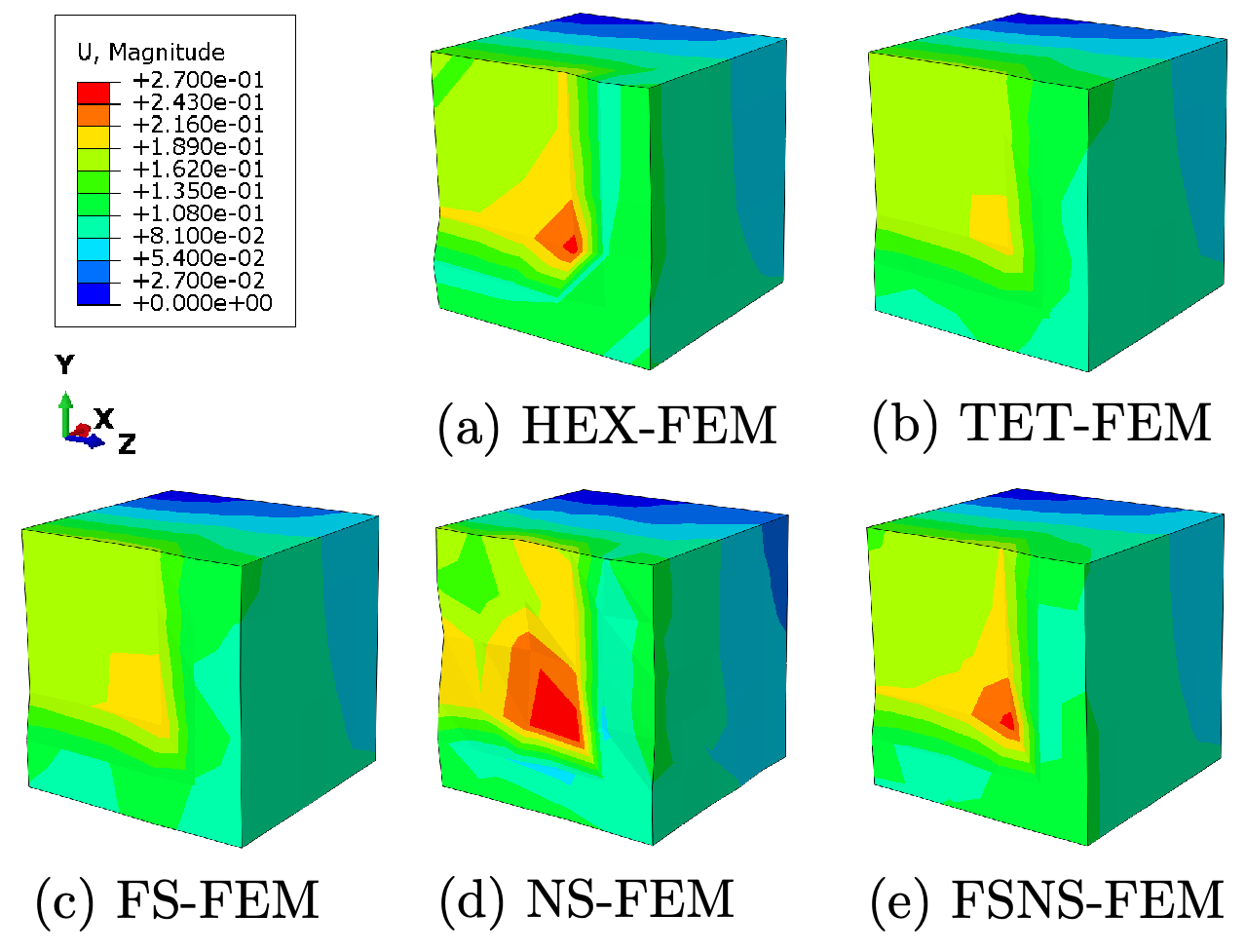}
	\caption{Snapshot taken at $t=50$ ms in Abaqus of the displacement magnitude $||\mathbf{u}||$ in mm for each numerical method.}      \label{fig:examples_die_res}                     
\end{figure}

\begin{figure}[h]
	\centering                              
	\includegraphics[scale=1]{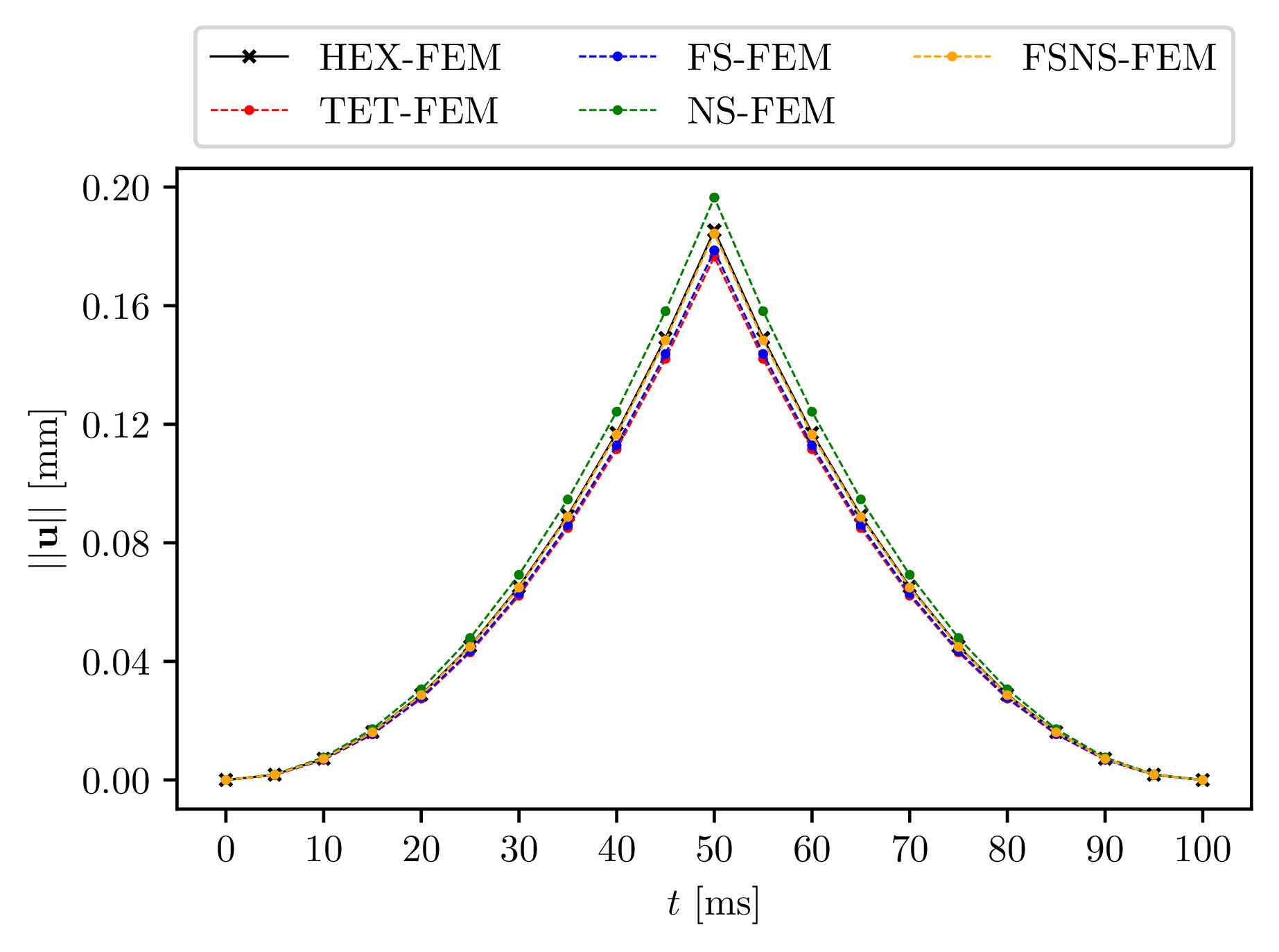}
	\caption{Averaged displacement magnitude $||\mathbf{u}||$ on excited surface $\mathcal{S}$ during the loading step $t \in [0,50]$ ms and unloading step $t \in [50,100]$ ms  for each method.}
	\label{fig:examples_die_plot}                     
\end{figure}

\noindent For the dielectric elastomer problem, we tested the implementations of the numerical methods on a simple electromechanically coupled cube with passive nearly incompressible Neo-Hookean material. The edge length was set to $l=10$ mm. We defined 256 nodes within the geometry which results into a discretisation of 1050 TET elements for FEM (TET-FEM), 2250 face-based SDs for FS-FEM and 256 node-based SDs for NS-FEM. For FSNS-FEM, 256 node-based SDs are over-layered on top of 2250 face-based SDs. The material-depend parameters necessary to describe the mechanical  and electrical behaviour are summarised in Table \ref{tab:examples_die_params}. 
\begin{table}[h!]
\centering
\begin{tabular}{c c c}
\hline
$\mu$ [kPa ]& $\lambda$ [kPa] & $\epsilon$ [Fmm$^{-2}$]\\
\hline
$2000$ & $666.67$ & $1$  
\end{tabular}
\caption{Material parameters necessary to describe the passive and active behaviour of the DEA cube.}
\label{tab:examples_die_params}
\end{table}
The cube was fixed in three directions preventing a rigid body motion and actuated by an external potential. The external potential was partially applied on the cube's front face and varied over time. We defined it as an electrical boundary condition $\overline{\varphi}(t)$. The potential on the opposite face was fixed at $\overline{\varphi}_0 = 0$ mV. The simulation setup is depicted in Figure \ref{fig:examples_die_bc}.
Here, $\mathcal{S}$ defines the excitation surface on which $\overline{\varphi}(t)$ is applied. The contraction simulation was carried out in two steps. First, $\overline{\varphi}(t)$ is linearly increased from $0$ mV to $100$ mV over a time span of $50$ ms. This lead to a compression of the cube. We denote the first step as the loading step. Second, $\overline{\varphi}(t)$ was linearly decreased back to $0$ mV over a time span of $50$ ms. The cube then returned back to its undeformed state. We denote the second step as unloading step. Additionally, a reference simulation on a hexahedral mesh (HEX) with 125 fully integrated elements was conducted to evaluate the accuracy of our implementations. We will denote the reference simulation as HEX-FEM. In Figure \ref{fig:examples_die_res}, snapshots taken in Abaqus at $t=50$ ms for each numerical method are depicted showing the nodal displacement magnitudes $||\mathbf{u}||$ in mm. The compression of the cube's front face is clearly visible with indentations around $\mathcal{S}$. 
To further compare the different solutions over time, we used a scalar representation of the deformation at each time.  Therefore, the nodal displacement solutions on $\mathcal{S}$ were averaged at specific times with a constant increment of $\Delta t=5$ ms. The resulting deformation curves are depicted in Figure \ref{fig:examples_die_plot}.
For each method, the curve starts at $0$ mm and monotonically increases until it reaches its amplitude at $t=50$ ms. Then, the curve decreases monotonically until the initial. We observe that the simulations using TET-FEM and FS-FEM yield deformations that are below the reference solution. This indicates an overestimated stiffness caused by volumetric locking for both methods. This phenomenon appears in linear TET meshes combined with nearly incompressible material behaviour according to \cite{Liu.2019}. The simulation using NS-FEM yields a deformation curve which is above the reference solution. This indicates an overestimated softness which corresponds to the limitations of NS-FEM stated in \cite{Wang.2015}. The simulation using FSNS-FEM returns results which are nearly identical to the reference solution indicating a high accuracy. For each S-FEM and TET-FEM we compute the relative error averaged over all $n_t$ time steps with 
\begin{equation}
\overline{e}_r = \frac{1}{n_t} \sum_{i=1}^{n_t} | 1-\frac{||u||_i}{||u||^{\text{ref}}_i} |
\label{eq:example_mean_err}
\end{equation}
where  $|u||_i$ is the averaged displacement magnitude at time step $i$ of the respective method depicted in Figure \ref{fig:examples_die_plot} and $||u||^{\text{ref}}_i$ is the reference obtained by HEX-FEM. The results are listed in Table \ref{tab:examples_die_er} together with the total central processing unit (CPU) times of each S-FEM and TET-FEM simulation. 
\begin{table}[h!]
	\centering
	\begin{tabular}{c  c c }
		\hline
		method & $\overline{e}_r$ [-] & CPU time [s] \\
		\hline
		TET-FEM & 0.047 & 299.9 \\
		FS-FEM & 0.036 & 39.3 \\
		NS-FEM & 0.061 & 11.0 \\
		FSNS-FEM & 0.005 & 61.7 \\
	\end{tabular}
	\caption{Mean relative error $\overline{e}_r$ over simulation time with respect to HEX-FEM and total CPU time for each numerical method.}
	\label{tab:examples_die_er}
\end{table}
The NS-FEM simulation runs the shortest with a CPU time of $11$ s but produces results with the highest error with $\overline{e}_r = 6.1 \%$. The remaining methods return sufficiently accurate solutions where the error is below $5 ~\%$. The simulation using TET-FEM runs the longest with a CPU time of around $300$ s which is nearly five times higher than the second longest simulation using FSNS-FEM. This result highlights the inefficiency of using TET-FEM compared to FS-FEM and FSNS-FEM. The FSNS-FEM simulation has the smallest error with $\overline{e}_r= 0.5  ~\%$ with a total CPU time of $61.7$ ms which is nearly five times shorter than for TET-FEM but still the second longest simulation. The FS-FEM simulation runs second shortest with a total CPU time of $39.3$ s and an error of $\overline{e}_r= 3.6  ~\%$. However, the observation in Figure \ref{fig:examples_die_plot} shows an overestimation of the stiffness over time when using FS-FEM. Therefore, we conclude that FSNS-FEM is the most suitable method to simulate the dielectric elastomer problem because it overcomes the overly-stiff behaviour of TET elements without overestimating the softness resulting in the most accurate solution with an error below $1 ~\%$ which is a sufficient trade-off for the longer CPU time.


\subsection{Coupled myocardial cube}
\label{subsec:examples_myo_cube}
\begin{figure}[h!]
	\centering                              
	\includegraphics[scale=0.2]{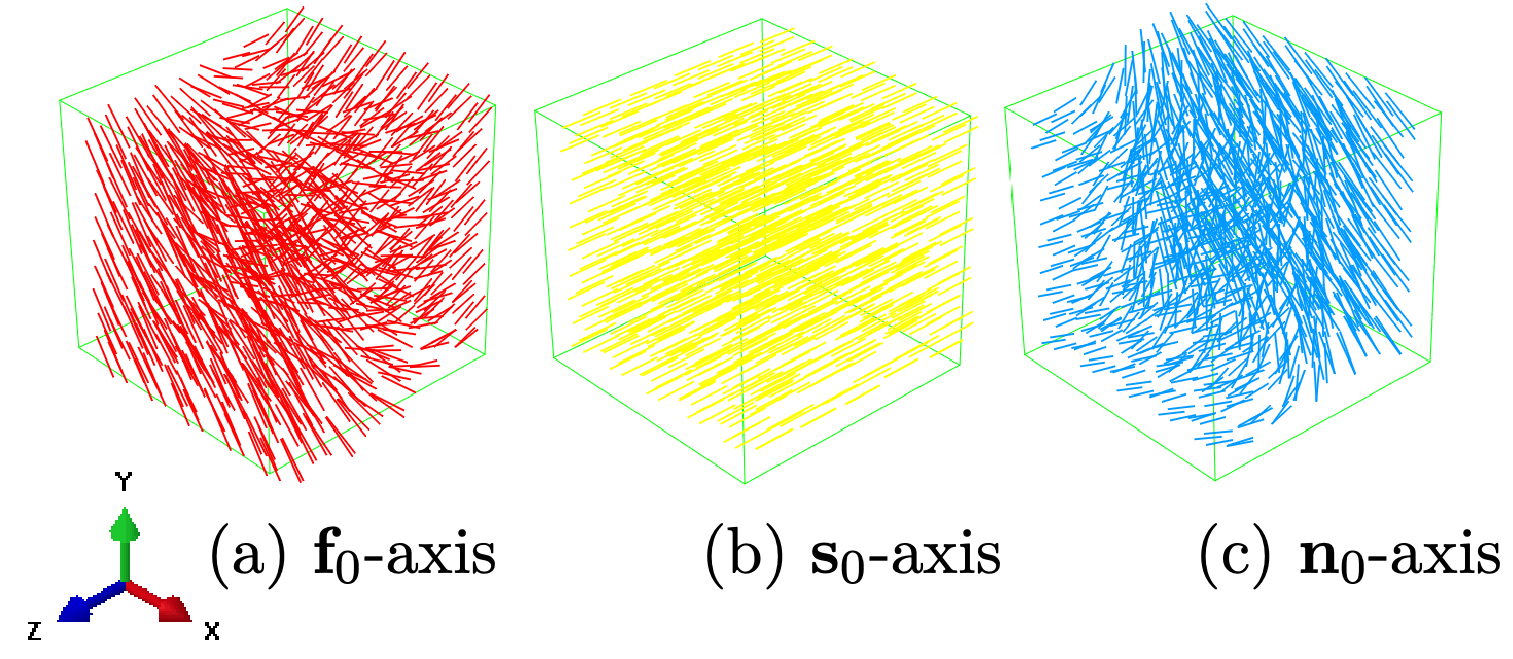}
	\caption{Material $\mathbf{f}_0$-, $\mathbf{s}_0$- and $\mathbf{n}_0$ - directions for each TET element. The fibre direction $\mathbf{f}_0$ rotates smoothly from $+60^\circ$ on the front face to $-60^\circ$ on the back face with respect to the global $y$-axis. The sheet direction $\mathbf{s}_0$ is aligned to the global $z$-axis for each element. The normal direction $\mathbf{n}_0$ results from the normalized cross-product between $\mathbf{f}_0$ and $\mathbf{s}_0$.}            
	\label{fig:examples_myo_orien}                     
\end{figure}

\begin{figure}[h!]
	\centering                              
	\centering                              
	\includegraphics[scale=0.2]{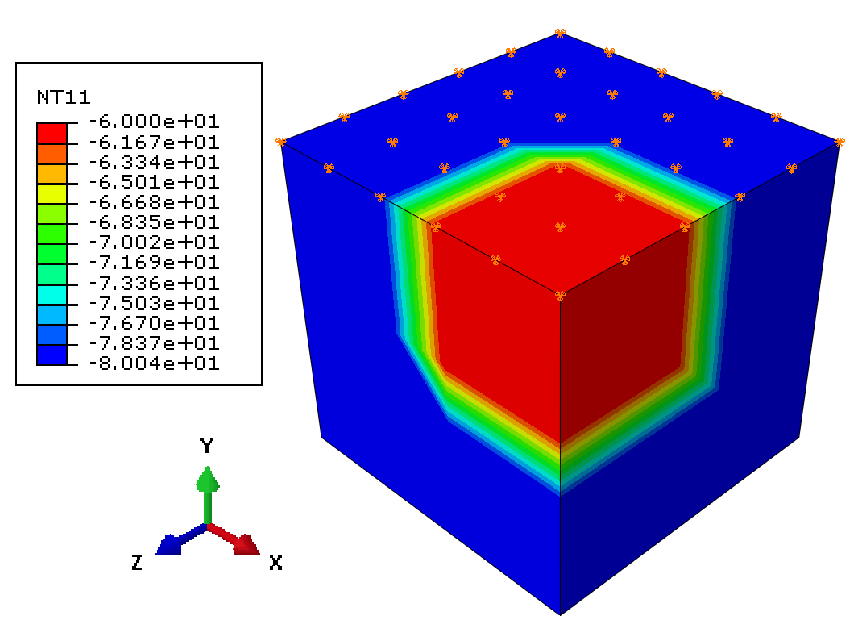}
	\caption{The nodes inside the activation region (red) are depolarised from $-80$ mV to $-60$ mV which generates an electric field inside the cube and triggers the generation of electric potential based on the Aliev-Paniflov model. The mechanical boundary conditions are visualised by the orange cones. The nodes on the top face at $(x, y=10,z)$ mm are fixed in all spatial directions.}            
	\label{fig:examples_myo_act}                     
\end{figure}

\begin{figure}[h!]
	\centering                              
	\includegraphics[scale=0.2]{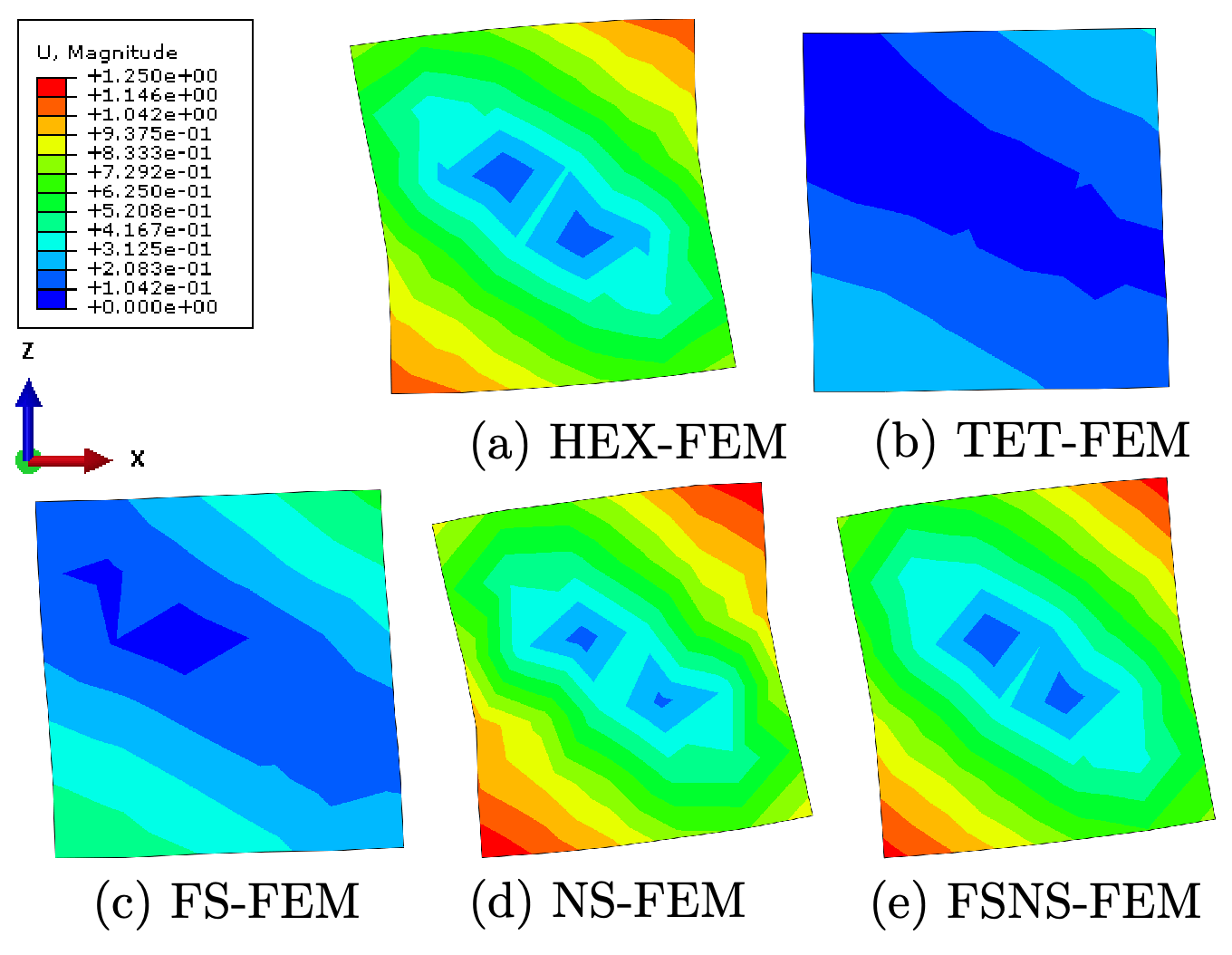}
	\caption{Snapshot taken at $t=75$ ms of the displacement magnitude $||u||$ in mm for each numerical method. We observe the bottom face of the cube at $(x,y=0,z)$ mm.}            
	\label{fig:examples_myo_res}                     
\end{figure}

\begin{figure}[h]
	\centering                                                       
	\includegraphics[scale=1]{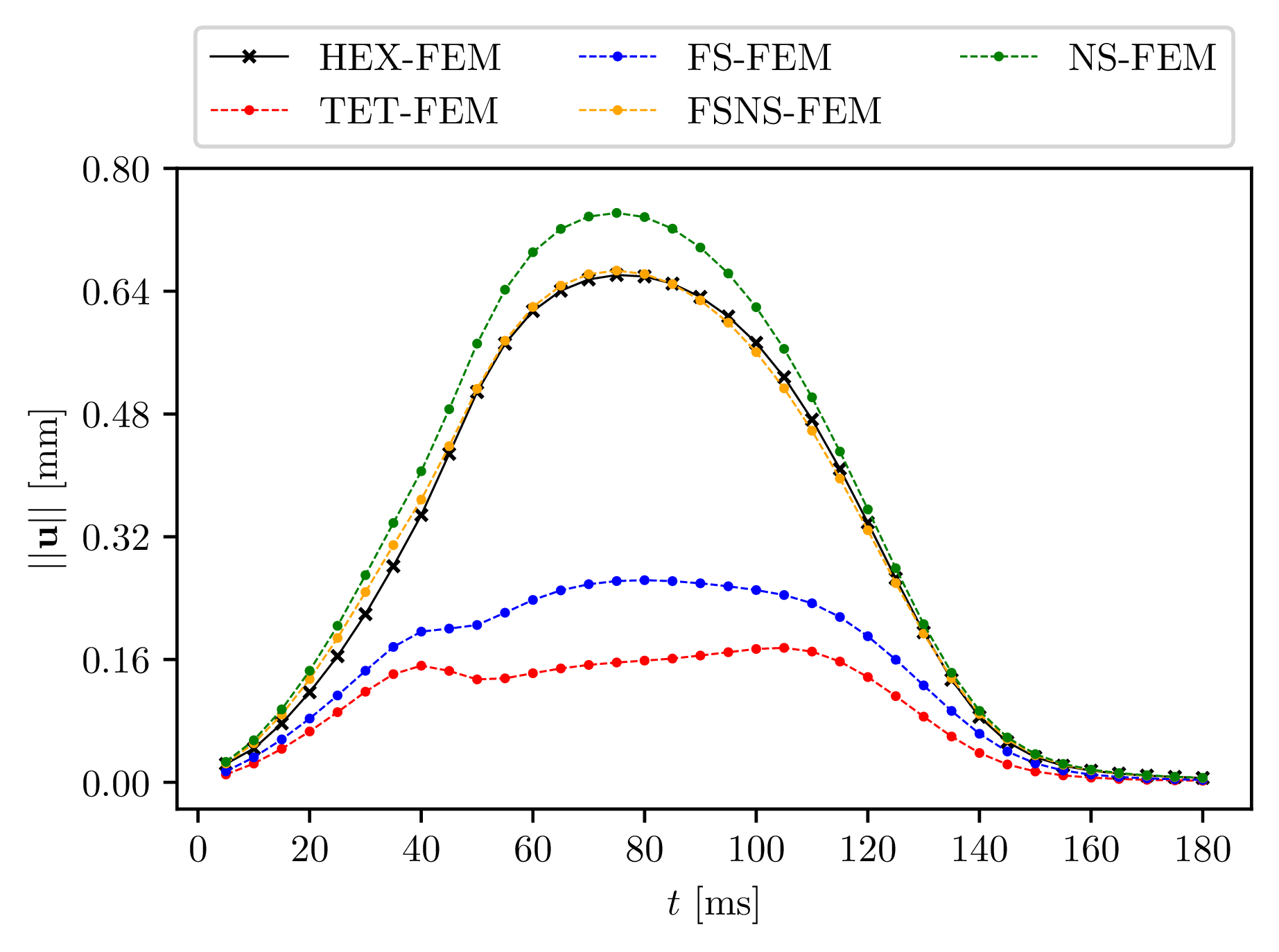}
	\caption{Averaged displacement magnitude $||\mathbf{u}||$ on bottom face at $(x,y=0,z)$ mm during the contraction step $t \in [0,180]$ ms for each method.} 
	\label{fig:examples_myo_plot}                     
\end{figure}

\noindent For this numerical example, the used geometry represents a section of the myocardium where the front face represents the sub-epicardium and the back face represents the sub-endocardium. The same cube model with an identical discretisation as for the dielectric problem was utilized. In summary, 256 nodes were defined, resulting in a mesh consisting of 1050 TET elements for TET-FEM, 2250 face-based SDs for FS-FEM, 256 node-based SDs for NS-FEM, and a combination of face-based/node-based SDs for FSNS-FEM. To mimic the orthotropic behaviour of the myocardium based on the Holzapfel-Ogden model, given in (\ref{eq:material_HO_energy}), we implemented rule-based local fibre, sheet and sheet-normal directions, $\mathbf{f}_0$, $\mathbf{s}_0$ and $\mathbf{n}_0$, respectively. Globally, the fibres were aligned by $+60^\circ$ on the sub-epicardium (front face in Figure \ref{fig:examples_myo_orien}.a) and $-60^\circ$ on the sub-endocardium (back face in Figure \ref{fig:examples_myo_orien}.a) with respect to the global $y$-axis. The rotation of the fibres throughout the myocardium is fairly smooth. For the TET mesh, we assigned a local $\mathbf{f}_0^e-\mathbf{s}_0^e-\mathbf{n}_0^e$ coordinate system to each element $e$. For S-FEM, the coordinate system was assigned to each SD. To compute the smoothed directional vectors $\mathbf{f}_0^k-\mathbf{s}_0^k-\mathbf{n}_0^k$, we adopted the rule from \cite{Martonova.2023}. There, the local coordinate system of the face-based or node-based SD is derived by averaging the $\mathbf{f}_0^e-\mathbf{s}_0^e-\mathbf{n}_0^e$ from each element connected to either face $k$ or node $k$ with
\begin{equation}
	\mathbf{f}_0^k = \frac{1}{n_k} \sum_{e=1}^{n_k} \mathbf{f}_0^e, ~ \mathbf{s}_0^k = \frac{1}{n_k} \sum_{e=1}^{n_k} \mathbf{s}_0^e
	~ \text{and} ~ \mathbf{n}_0^k = \frac{1}{n_k} \sum_{e=1}^{n_k} \mathbf{n}_0^e.
	\label{eq:examples_myo_SD_orien}
\end{equation}
In Figure \ref{fig:examples_myo_orien}, the material $\mathbf{f}_0$-, $\mathbf{s}_0$ and $\mathbf{n}_0$ - axes are depicted for each TET element. There, the front face at $(x,y,z=0)$ mm represents the sub-epicardium and the back face at $(x,y,z=10)$ mm represent the sub-endocardium.
To compute the passive stress in (\ref{eq:material_stress_myo}), we adopted the parameters given in \cite{Martonova.2021}. There, the strain-energy from (\ref{eq:material_HO_energy}) was fitted to experimental data obtained from an uniaxial extension test and simple shear test conducted simultaneously.
\begin{table}[h!]
\centering
\begin{tabular}{c c c c c c c c c}
\hline
$\kappa$ [kPa] &  $a$ [kPa] & $b$ [-]  &$a_f$ [kPa]  & $b_f$ [-]  & $a_s$ [kPa]  & $b_s$ [-]  & $a_{fs}$ [kPa]  & $b_{fs}$ [-]   \\ 
 \hline
1000 &1.665& 1.237& 7.822 & 0.008 & 0 & 0 & 1.342 & 9.178
\end{tabular}
\caption{Material parameters necessary to describe the passive behaviour of the myocardial cube based on the Holzapfel-Ogden model.}
\label{tab:eaxmples_myo_pas}
\end{table}
For the calculation of the active stress in (\ref{eq:material_stress_myo}), we used the parameters summarised in Table \ref{tab:example_myo_act}. There, the values were chosen to achieve stable simulations for each method. 
\begin{table}[h!]
\centering
\begin{tabular}{c c c c c c}
\hline
$k_T$ [kPamV$^{-1}$] & $a_0$ [mV$^{-1}$] & $a_\infty$ [mV$^{-1}$] & $\xi$ [mV$^{-1}$] & $\varphi_r$ [mV] & $\overline{\varphi}$ [mV]  \\
\hline
0.005 & 1 & 0.1 & 0.1 & -80 & -80
\end{tabular}
\caption{Material parameters necessary to describe the active behaviour of the myocardial cube.}
\label{tab:example_myo_act}
\end{table}
To account for the electrical flux $\mathbf{q}$, the scalar conductivity from(\ref{eq:material_D}) was set to $d=0.01$ mm$^2$ms$^{-1}$. For simplistic reasons, we assumed that the resulting orthotropic material behaviour does not affect the spatial propagation of the electric potential $\varphi$. Therefore, $\mathbf{q}$ is independent of the fibre-orientations. The coupled contraction simulation was run in three steps. In the initial step, each node inside the cube were polarised at $-80$ mV for $1$ ms. In the next step, the nodal potentials in a specified region were increased to $-60$ mV  for $1$ ms. The used activation node-set was constructed to enable a stable simulation for each method. The chosen activation region together with the mechanical boundary conditions for the simulation are shown in Figure \ref{fig:examples_myo_act}. 
This activation step triggers the generation of electric potential and an electric Field. As shown in Table \ref{tab:numerics_overview_myo}, the electric potential is generated on the TET mesh for each method. The electric field is either generated on the TET mesh or the SD mesh. In the last step, all electric boundary conditions were omitted. The source term $I^\varphi$ governs the generation and temporal evolution of the electric potential and the electric field $\mathbf{E}$ governs the spatial propagation of the electric potential which results in the deformation of the cube. The contraction step was defined for $180$ ms. For all steps, the nodes at the cube's top were fixed in each direction to prevent translation. Similar to the dielectric problem, a reference simulation on a hexahedral (HEX) mesh with 125 fully integrated elements was conducted to evaluate the accuracy of the numerical methods. In Figure \ref{fig:examples_myo_res}, snapshots taken in Abaqus at $t=75$ ms for each numerical method are depicted showing the nodal displacement magnitude $||\mathbf{u}||$ in mm on the cube's bottom at $(x,y=0,z)$ mm.
The observed deformation modes indicate compression and torsion of the cube for each method. The largest displacement appears in the outer corner, while the centre remains nearly static. Similar to the dielectric problem, we used a scalar representation of the deformation to compare each method. Therefore, the nodal displacement magnitudes at the cube's bottom were averaged at specific times with a constant increment of $\Delta t=5$ ms. The resulting deformation curves for the contraction step are depicted in Figure \ref{fig:examples_myo_plot}. 
The reference solution curve and the curves obtained from NS-FEM and FSNS-FEM increase until their respective amplitudes are reached at $t=75$ ms and then decrease back to the undeformed state. Observing the results from TET-FEM and FS-FEM, the curves first increase until $t=40$ ms and the decrease for two time steps. This valleys in combination with the stiffer behaviour indicate the appearance of volumetric locking, similar to the dielectric problem. Using FS-FEM reduces the volumetric locking but the material behaviour is still considerably stiff. Using NS-FEM, the deformation is larger than the reference which indicates an overestimated softness, similar to the observation from the DEA problem. The results obtained from the FSNS-simulation are in best agreement with the reference solutions. For each S-FEMs and TET-FEM we again compute $\overline{e}_r$ using (\ref{eq:example_mean_err}). The results are listed in Table \ref{tab:examples_myo_er} together with the total CPU time for each method. 
\begin{table}
	\centering
	\begin{tabular}{c  c c }
		\hline
		method & $\overline{e}_r$ [-] & CPU time [s] \\
		\hline
		TET-FEM & 0.686& 614.4 \\
		FS-FEM & 0.533 & 72.9 \\
		NS-FEM & 0.115 & 43.9\\
		FSNS-FEM & 0.028& 228.0 \\
	\end{tabular}
	\caption{Mean relative error $\overline{e}_r$ over simulation time with respect to the HEX-FEM solution and total CPU time for each numerical method.}
	\label{tab:examples_myo_er}
\end{table}
The use of TET-FEM results in the largest error with $\overline{e}_r = 68.6~ \%$ and takes the longest with a CPU time of $614.4$ s. Similar to the DEA example, this method tends to be relatively ineffective. The simulation using NS-FEM runs the fastest and produces the second smallest error with $\overline{e}_r = 11.5~ \%$. The use of FS-FEM results in the second-fastest simulation with a CPU time of $72.9$ s. Nevertheless, an error of $\overline{e}_r = 53.3 ~\%$ renders this method still ineffective. The solution obtained from the FSNS-FEM simulation shows the best approximation with an error of $\overline{e}_r = 2.8 ~\%$. Relative to FS-FEM and NS-FEM, FSNS-FEM still requires a longer CPU time with $228$ s. Nevertheless, we still conclude FSNS-FEM to be the most suitable method to simulate the myocardial tissue because it overcomes the volumetric locking with a small enough error which compensates for the longer CPU time.


\section{Concluding remarks}
\label{sec:concluding remarks}

\section*{Conclusion}
In this work, we used linear FEM and various S-FEMs to numerically approximate and solve the electromechanically coupled problem for a dielectric elastomer and myocardial tissue. The commercial software Abaqus served as the simulation environment, with electromechanically coupled UELs implemented using TET-FEM, FS-FEM, NS-FEM, and FSNS-FEM. Two numerical examples were simulated: the electrically induced contraction of a DEA and a myocardial tissue sample. \\ For both cases, a cube served as the benchmark geometry. In the DEA example, passive material behaviour was modelled by the nearly incompressible Neo-Hookean energy function, and active behaviour by the ideal dielectric elastomer energy function. Contraction was induced through a temporally varying external potential, with a HEX mesh reference solution used for comparison. TET-FEM and FS-FEM overestimated stiffness due to volumetric locking, NS-FEM overestimated softness, while FSNS-FEM delivered the closest match to the reference solution, achieving high accuracy at relatively low computational cost thanks to the use of linear shape functions.  \\
In the myocardium example, the nearly incompressible Holzapfel--Ogden energy function was used to model passive behaviour, with orthotropy incorporated via local fibre directions. Contraction was triggered via depolarising a certain node set. The electric field differed between methods due to gradient smoothing effects, influencing the resulting contraction. Comparisons with the HEX mesh again showed high volumetric locking for TET-FEM, reduced locking for S-FEMs, persistent stiffness/softness tendencies for FS-FEM and NS-FEM, and the highest accuracy for FSNS-FEM.  \\
While FSNS-FEM proved to be the most accurate method for both benchmark problems, the simplified cube geometries and boundary conditions limit the physiological realism of the current simulations. Future work incorporating anatomically realistic geometries, refined boundary conditions, and more advanced constitutive models could improve predictive capacity.  \\
Overall, FSNS-FEM emerged as a robust and reliable approach for simulating the electromechanical behaviour of both dielectric elastomers and myocardial tissue, consistently achieving high accuracy at relatively low computational cost. These results underscore its strong potential for more realistic and clinically relevant electromechanical simulations, paving the way for advanced tissue modelling and patient-specific computational studies.

\section*{Acknowledgements} 
\noindent This work is funded by the Deutsche Forschungsgemeinschaft (DFG, German Research Foundation) project no.~496647562. 
\section*{Declaration of competing interests}  
\noindent There is no conflict of interest to declare. 
\section*{Supplementary material}  
\noindent After publication of the final version of the manuscript, an implementation of all the user element subroutines will be available at our Github repository \\ \text{https://github.com/TanTran1512/SFEM_electromechanics}.

\appendix

\section{Linearisation of the local residuals}
\noindent
To numerically solve the non-linear coupled problem given by $\mathbf{R}^a = (\mathbf{R}^a_\mathbf{u}, R^a_\varphi)^T=\mathbf{0}$, the Newton-Raphson scheme was employed for this work. The necessary linearisation of $\mathbf{R}^a $ results in the following tangent stiffness matrices 
\begin{subequations}
    \begin{align}
	&\mathbf{K}^{ab}_\mathbf{uu} = -\frac{\partial \mathbf{R}^a_\mathbf{u}}{\partial \mathbf{u}^b}, ~
	\mathbf{K}^{ab}_{\mathbf{u}\varphi}= -\frac{\partial \mathbf{R}^a_\mathbf{u}}{\partial \varphi^b},\\
	&\mathbf{K}^{ab}_{\varphi \mathbf{u}} = -\frac{\partial R^a_\varphi}{\partial \mathbf{u}^b} ~ \text{and} ~
	\mathbf{K}^{ab}_{\varphi \varphi}= -\frac{\partial R^a_\varphi}{\partial \varphi^b}.
\end{align}
\label{eq:appendix_K_all}
\end{subequations}
Assembling the individual tangents leads to the local tangent stiffness matrix $\mathbf{K}^{ab}$. The computation of the tangents differs between the coupled problem formulations of the dielectric elastomer and the myocardium. To implement the numerical scheme for the coupled dielectric elastomer, we adopt the formulations given in \cite{Henann.2013}. For the coupled myocardial problem, we adopt the formulations given in \cite{Goktepe.2010}. In both works, the governing equations are given for FEM. The procedure is equivalent for S-FEM. In the following, we only state the tangent expressions for S-FEM.


\subsection{Dielectric elastomer formulation}
\label{subsec:appendix_die}
\noindent 
Linearising the mechanical residual $\mathbf{R}^a_\mathbf{u}$ from (\ref{eq:numerics_Ru}) with respect to $\mathbf{u}$, results into the following explicit expression of $\mathbf{K}^{ab}_\mathbf{uu}$
\begin{equation}
	\mathbf{K}^{ab}_\mathbf{uu} = \int_{\mathcal{B}^k} \nabla N^a :\mathbb{C}_{\mathbf{uu}} : \nabla N^b dv +  \int_{\mathcal{B}^k}  (\nabla N^a \bm{\sigma}  \nabla N^b) \mathbf{I} dv
	\label{eq:appendix_die_Kuu}
\end{equation}
There, the first part describes the material non-linearity governed by the fourth-order spatial elasticity tensor $\mathbb{C}_{\mathbf{uu}}$.  According to \cite{Federico.2012}, $\mathbb{C}_{\mathbf{uu}}$ is defined by the push-forward of the material elasticity tensor with
\begin{equation}
	\mathbb{C}_{\mathbf{uu}} = \frac{1}{J} (\mathbf{F}^k \otimes \mathbf{F}^k) : 2 \frac{\partial \mathbf{S}}{\partial \mathbf{C}^k} : (\mathbf{F}^k \otimes \mathbf{F}^k)^T
	\label{eq:appendix_die_Cuu}
\end{equation}
where $\mathbf{S} = $$J(\mathbf{F}^k)^{-1} \bm{\sigma} (\mathbf{F}^k)^{-T}$ denotes the second Piola-Kirchhoff stress tensor. The second part describes the geometric non-linearity governed by the Cauchy stress $\bm{\sigma}$ from (\ref{eq:material_stress_dielectric}) consisting of the passive part and active part. Evaluating the remaining tangents, results in the following expressions
\begin{align}
	&\mathbf{K}^{ab}_{\mathbf{u} \varphi} =  \int_{\mathcal{B}^k} \nabla N^a :\mathbb{C}_{\mathbf{u} \varphi}: \nabla N^b  dv , \label{eq:appendix_Kuphi_die} \\
	&\mathbf{K}^{ab}_{\varphi \mathbf{u}} =-\int_{\mathcal{B}^k} \nabla N^a : \mathbb{C}_{\varphi \mathbf{u}}:  \nabla N^b dv  ~ \text{and} ~
	\label{eq:appendix_Kphiu_die} \\
	&\mathbf{K}^{ab}_{\varphi \varphi} =- \int_{\mathcal{B}^k} \nabla N^a \mathbb{C}_{\varphi \varphi} \nabla N^b dv. \label{eq:appendix_Kphiphi_die}
\end{align}	
There, the sensitivities $\mathbb{C}_{\mathbf{u} \varphi}$, $ \mathbb{C}_{\varphi \mathbf{u}} $ and $\mathbb{C}_{\varphi \varphi}$ are formulated in terms of partial derivatives. We adopt the definitions given in \cite{Henann.2013}
\begin{equation}
	\mathbb{C}_{\mathbf{u} \varphi} = -\frac{\partial \bm{\sigma}}{\partial \mathbf{E}^k}, ~  \mathbb{C}_{\varphi \mathbf{u}} = \frac{\partial \mathbf{D}}{\partial \mathbf{F}^k} \mathbf{F}^k ~ \text{and} ~
	\mathbb{C}_{\varphi \varphi} = -\frac{\partial \mathbf{D}}{\partial \mathbf{E}^k}.
\end{equation}
An explicit calculation of the above expressions can be found in our code implementations.


\subsection{Myocardium formulation}
\label{subsec:appendix_myo}
\noindent The linearisation of the mechanical residual for the coupled myocardial problem, results into the same form from (\ref{eq:appendix_die_Kuu}). There, the spatial elasticity tensor $\mathbb{C}_{\mathbf{uu}}$ is computed from the passive Holzapfel-Ogden model. According to \cite{Goktepe.2010}, we have
\begin{align}
	\mathbb{C}_\mathbf{uu} &=\kappa J (2J-1) \mathbf{I} \otimes \mathbf{I} - 2 \kappa (J^2-1) \mathbb{I}^\text{sym}   \nonumber \\
	& +\mathbb{P}:[\overline{\mathbb{C}} + \frac{2}{3}(J\overline{\bm{\sigma}}:\mathbf{I}) \mathbb{I}^\text{sym}]:\mathbb{P}
	-\frac{2}{3} (\mathbb{P} :J\bm{\sigma} \otimes \mathbf{I} + \mathbf{I}:J\overline{\bm{\sigma}} \otimes \mathbb{P})
	\label{eq:appendix_my_Cuu}
\end{align}
with the modified Cauchy stress $\overline{\bm{\sigma}}$ from (\ref{eq:material_HO_modified_stress}) and a modified elasticity tensor $\overline{\mathbb{C}}$. The explicit form is also given in \cite{Goktepe.2010} by
\begin{align}
	\overline{\mathbb{C}} &= 4 \frac{\partial\Psi^{\text{iso}}}{\partial \overline{I}_{\text{iso}}} (\overline{\mathbf{b}}^k \otimes \overline{\mathbf{b}}^k) 
									   + 4 \frac{\partial\Psi^{\text{f}}}{\partial \overline{I}_f} (\overline{\mathbf{f}}^k \otimes \overline{\mathbf{f}}^k \otimes \overline{\mathbf{f}}^k \otimes \overline{\mathbf{f}}^k) \nonumber \\
									   &+  4 \frac{\partial\Psi^{\text{s}}}{\partial \overline{I}_s} (\overline{\mathbf{s}}^k \otimes \overline{\mathbf{s}}^k \otimes \overline{\mathbf{s}}^k \otimes \overline{\mathbf{s}}^k) \nonumber \\
									   &+ \frac{\partial\Psi^{\text{fs}}}{\partial \overline{I}_{fs}} 	(\overline{\mathbf{f}}^k \otimes \overline{\mathbf{s}}^k + \overline{\mathbf{s}}^k \otimes \overline{\mathbf{f}}^k) \otimes 
									   (\overline{\mathbf{f}}^k \otimes \overline{\mathbf{s}}^k + \overline{\mathbf{s}}^k \otimes \overline{\mathbf{f}}^k)
\end{align}
 in terms of the isochoric left Cauchy-Green tensor from (\ref{eq:fundamentals_strain_iso}), the scalar derivatives of the isochoric invariants from (\ref{eq:material_HO_invariants}) and the smoothed isochoric directions from (\ref{eq:examples_myo_SD_orien}). Evaluating the remaining tangents, results in the following expressions
\begin{align}
	&\mathbf{K}^{ab}_{\mathbf{u} \varphi} =  \int_{\mathcal{B}^k} (\nabla N^a \mathbb{C}_{\mathbf{u} \varphi})  N^b  dv , \label{eq:appendix_Kuphi_myo} \\
	& \mathbf{K}^{ab}_{\varphi \mathbf{u}} =-\int_{\mathcal{B}^k} \nabla N^a :\mathbb{C}_{\varphi \mathbf{u}}:\nabla N^b dv  ~ \text{and} ~
	\label{eq:appendix_Kphiu_myo} \\
	&\mathbf{K}^{ab}_{\varphi \varphi} = \int_{\mathcal{B}^k}[-N^a \frac{1}{\Delta t} N^b - \nabla N^a \mathbb{C}_{\varphi \varphi} \nabla N^b  + N^a \frac{\partial I^{\varphi}}{\partial \varphi^k} N^b]dv \label{eq:appendix_Kphiphi_myo}.
\end{align}	
The sensitivities are computed based on the following partial derivatives
\begin{equation}
	\mathbb{C}_{\mathbf{u} \varphi} = \frac{\partial \bm{\sigma}}{\partial \varphi^k}, ~  \mathbb{C}_{\varphi \mathbf{u}} = \frac{\partial \mathbf{q}}{\partial \mathbf{F}^k} \mathbf{F}^k ~ \text{and} ~
	\mathbb{C}_{\varphi \varphi} = -\frac{\partial \mathbf{q}}{\partial \mathbf{E}^k}.
\end{equation}
To obtain $\mathbb{C}_{\mathbf{u} \varphi}$, the definition in \cite{Goktepe.2010} was used. Due to the similar formulation between $\mathbf{D}$ and $\mathbf{q}$, the implementation of $\mathbb{C}_{\varphi \mathbf{u}}$ and $\mathbb{C}_{\varphi \varphi}$ was based on \cite{Henann.2013}.
As mentioned previously, the source term is computed on the TET mesh instead of the SD mesh. Therefore, the tangent stiffness $\mathbf{K}^{ab}_{\varphi \varphi}$ for S-FEM implemented with the terms $\frac{1}{\Delta t} = 0$ and  $\frac{\partial I^{\varphi}}{\partial \varphi^k}=0$.  The temporal evolution of $\varphi$ is computed inside an additional user subroutine called HETVAL. This subroutine is originally used to define heat generation inside an element. In our work it is used to define the generation and evolution over time of $\varphi$ inside the TET element. An explicit computation of the sensitivities and resulting tangent stiffness matrices can be found in our code implementations. 

\bibliographystyle{apalike}
\bibliography{literature}

\begin{thebibliography}{}

\bibitem[Aliev and Panfilov, 1996]{Aliev.1996}
Aliev, R.~R. and Panfilov, A.~V. (1996).
\newblock A simple two-variable model of cardiac excitation.
\newblock {\em Chaos, Solitons {\&} Fractals}, 7(3):293--301.

\bibitem[Bonet and Wood, 1997]{Bonet.1997}
Bonet, J. and Wood, R.~D. (1997).
\newblock {\em Nonlinear continuum mechanics for finite element analysis}.
\newblock {Cambridge Univ. Press}, Cambridge.

\bibitem[Cai and Zhou, 2019]{Cai.2019}
Cai, B. and Zhou, L. (2019).
\newblock A coupling electromechanical inhomogeneous cell-based smoothed finite
  element method for dynamic analysis of functionally graded piezoelectric
  beams.
\newblock {\em Advances in Materials Science and Engineering}, 2019(1).

\bibitem[{Dassault Systems}, 2016]{DassaultSystems.2016}
{Dassault Systems} (2016).
\newblock Abaqus online documentation.

\bibitem[Eriksson et~al., 2013]{Eriksson.2013}
Eriksson, T. S.~E., Prassl, A.~J., Plank, G., and Holzapfel, G.~A. (2013).
\newblock Modeling the dispersion in electromechanically coupled myocardium.
\newblock {\em International Journal for Numerical Methods in Biomedical
  Engineering}, 29(11):1267--1284.

\bibitem[Federico, 2012]{Federico.2012}
Federico, S. (2012).
\newblock Covariant formulation of the tensor algebra of non-linear elasticity.
\newblock {\em International Journal of Non-Linear Mechanics}, 47(2):273--284.
\newblock Nonlinear Continuum Theories.

\bibitem[G{\"o}ktepe et~al., 2011]{Goktepe.2011}
G{\"o}ktepe, S., Acharya, S. N.~S., Wong, J., and Kuhl, E. (2011).
\newblock Computational modeling of passive myocardium.
\newblock {\em International Journal for Numerical Methods in Biomedical
  Engineering}, 27(1):1--12.

\bibitem[G{\"o}ktepe and Kuhl, 2009]{Goktepe.2009}
G{\"o}ktepe, S. and Kuhl, E. (2009).
\newblock Computational modeling of cardiac electrophysiology: A novel finite
  element approach.
\newblock {\em International Journal for Numerical Methods in Engineering},
  79(2):156--178.

\bibitem[G{\"o}ktepe and Kuhl, 2010]{Goktepe.2010}
G{\"o}ktepe, S. and Kuhl, E. (2010).
\newblock Electromechanics of the heart: a unified approach to the strongly
  coupled excitation--contraction problem.
\newblock {\em Computational Mechanics}, 45(2-3):227--243.

\bibitem[Henann et~al., 2013]{Henann.2013}
Henann, D.~L., Chester, S.~A., and Bertoldi, K. (2013).
\newblock Modeling of dielectric elastomers: Design of actuators and energy
  harvesting devices.
\newblock {\em Journal of the Mechanics and Physics of Solids},
  61(10):2047--2066.

\bibitem[Holzapfel et~al., 2000]{Holzapfel.2000}
Holzapfel, G.~A., Gasser, T.~C., and Ogden, R.~W. (2000).
\newblock A new constitutive framework for arterial wall mechanics and a
  comparative study of material models.
\newblock {\em Journal of Elasticity}, 61(1/3):1--48.

\bibitem[Holzapfel and Ogden, 2009]{Holzapfel.2009}
Holzapfel, G.~A. and Ogden, R.~W. (2009).
\newblock Constitutive modelling of passive myocardium: a structurally based
  framework for material characterization.
\newblock {\em Philosophical transactions. Series A, Mathematical, physical,
  and engineering sciences}, 367(1902):3445--3475.

\bibitem[Jiang et~al., 2014]{Jiang.2014}
Jiang, C., Zhang, Z.-Q., Han, X., and Liu, G.-R. (2014).
\newblock Selective smoothed finite element methods for extremely large
  deformation of anisotropic incompressible bio--tissues.
\newblock {\em International Journal for Numerical Methods in Engineering},
  99(8):587--610.

\bibitem[Jiang et~al., 2015]{Jiang.2015}
Jiang, C., Zhang, Z.-Q., Liu, G.~R., Han, X., and Zeng, W. (2015).
\newblock An edge-based/node-based selective smoothed finite element method
  using tetrahedrons for cardiovascular tissues.
\newblock {\em Engineering Analysis with Boundary Elements}, 59:62--77.

\bibitem[Liu, 2019]{Liu.2019}
Liu, G.-R. (2019).
\newblock The smoothed finite element method (s-fem): A framework for the
  design of numerical models for desired solutions.
\newblock {\em Frontiers of Structural and Civil Engineering}, 13(2):456--477.

\bibitem[Liu et~al., 2007]{Liu.2007}
Liu, G.~R., Dai, K.~Y., and Nguyen, T.~T. (2007).
\newblock A smoothed finite element method for mechanics problems.
\newblock {\em Computational Mechanics}, 39(6):859--877.

\bibitem[Liu and Nguyen, 2010]{Liu.2010}
Liu, G.~R. and Nguyen, T.~T. (2010).
\newblock {\em Smoothed finite element methods}.
\newblock CRC Press, 1st ed. edition.

\bibitem[Liu et~al., 2009]{Liu.2009}
Liu, G.~R., Nguyen-Thoi, T., Nguyen-Xuan, H., and Lam, K.~Y. (2009).
\newblock A node-based smoothed finite element method (ns-fem) for upper bound
  solutions to solid mechanics problems.
\newblock {\em Computers {\&} Structures}, 87(1-2):14--26.

\bibitem[Martonova et~al., 2021]{Martonova.2021}
Martonova, D., Alkassar, M., Seufert, J., Holz, D., Duong, M.~T., Reischl, B.,
  Friedrich, O., and Leyendecker, S. (2021).
\newblock Passive mechanical properties in healthy and infarcted rat left
  ventricle characterised via a mixture model.
\newblock {\em Journal of the Mechanical Behavior of Biomedical Materials},
  119:104430.

\bibitem[Martonov{\'a} et~al., 2021]{Martonova.2021b}
Martonov{\'a}, D., Holz, D., Duong, M.~T., and Leyendecker, S. (2021).
\newblock Towards the simulation of active cardiac mechanics using a smoothed
  finite element method.
\newblock {\em Journal of biomechanics}, 115:110153.

\bibitem[Martonov{\'a} et~al., 2023]{Martonova.2023}
Martonov{\'a}, D., Holz, D., Duong, M.~T., and Leyendecker, S. (2023).
\newblock Smoothed finite element methods in simulation of active contraction
  of myocardial tissue samples.
\newblock {\em Journal of biomechanics}, 157:111691.

\bibitem[Mendizabal et~al., 2017]{Mendizabal.2017}
Mendizabal, A., {Bessard Duparc}, R., Bui, H.~P., Paulus, C.~J., Peterlik, I.,
  and Cotin, S. (2017).
\newblock Face-based smoothed finite element method for real-time simulation of
  soft tissue.
\newblock {\em Medical Imaging 2017: Image-Guided Procedures, Robotic
  Interventions, and Modeling}.

\bibitem[{Minh Tuan Duong}, 2014]{Minh.2014}
{Minh Tuan Duong} (2014).
\newblock {\em Hyperelastic Modeling and Soft-Tissue Growth Integrated with the
  Smoothed Finite Element Method-SFEM}.
\newblock Dissertation, {Rheinisch-Westf{\"a}lische Technische Hochschule
  Aachen}, Aachen.

\bibitem[Nash and Panfilov, 2004]{Nash.2004}
Nash, M.~P. and Panfilov, A.~V. (2004).
\newblock Electromechanical model of excitable tissue to study reentrant
  cardiac arrhythmias.
\newblock {\em Progress in biophysics and molecular biology}, 85(2-3):501--522.

\bibitem[Nguyen-Thoi et~al., 2009]{NguyenThoi.2009b}
Nguyen-Thoi, T., Liu, G.~R., Lam, K.~Y., and Zhang, G.~Y. (2009).
\newblock A face--based smoothed finite element method (fs--fem) for 3d linear
  and geometrically non--linear solid mechanics problems using 4--node
  tetrahedral elements.
\newblock {\em International Journal for Numerical Methods in Engineering},
  78(3):324--353.

\bibitem[Nguyen-Thoi et~al., 2010]{NguyenThoi.2010}
Nguyen-Thoi, T., Vu-Do, H.~C., Rabczuk, T., and Nguyen-Xuan, H. (2010).
\newblock A node-based smoothed finite element method (ns-fem) for upper bound
  solution to visco-elastoplastic analyses of solids using triangular and
  tetrahedral meshes.
\newblock {\em Computer Methods in Applied Mechanics and Engineering},
  199(45-48):3005--3027.

\bibitem[Townsend et~al., 2022]{Townsend.2022}
Townsend, N., Kazakiewicz, D., {Lucy Wright}, F., Timmis, A., Huculeci, R.,
  Torbica, A., Gale, C.~P., Achenbach, S., Weidinger, F., and Vardas, P.
  (2022).
\newblock Epidemiology of cardiovascular disease in europe.
\newblock {\em Nature reviews. Cardiology}, 19(2):133--143.

\bibitem[Wang et~al., 2015]{Wang.2015}
Wang, G., Cui, X.~Y., Feng, H., and Li, G.~Y. (2015).
\newblock A stable node-based smoothed finite element method for acoustic
  problems.
\newblock {\em Computer Methods in Applied Mechanics and Engineering},
  297:348--370.

\bibitem[Wang et~al., 2016]{Wang.2016}
Wang, S., Decker, M., Henann, D.~L., and Chester, S.~A. (2016).
\newblock Modeling of dielectric viscoelastomers with application to
  electromechanical instabilities.
\newblock {\em Journal of the Mechanics and Physics of Solids}, 95:213--229.

\bibitem[{World Health Organization}, 2021]{WHO.2021}
{World Health Organization} (2021).
\newblock Cardiovascular diseases (cvds).

\bibitem[Wriggers, 2008]{Wriggers.2008}
Wriggers, P. (2008).
\newblock {\em Nonlinear Finite Element Methods}.
\newblock {Springer Berlin Heidelberg}, Berlin, Heidelberg.

\bibitem[Zhao et~al., 2007]{Zhao.2007}
Zhao, X., Hong, W., and Suo, Z. (2007).
\newblock Electromechanical hysteresis and coexistent states in dielectric
  elastomers.
\newblock {\em Physical Review B}, 76(13).

\bibitem[Zheng et~al., 2019]{Zheng.2019}
Zheng, J., Duan, Z., and Zhou, L. (2019).
\newblock A coupling electromechanical cell-based smoothed finite element
  method based on micromechanics for dynamic characteristics of piezoelectric
  composite materials.
\newblock {\em Advances in Materials Science and Engineering}, 2019:1--16.

\end{thebibliography}

\end{document}